\pdfoutput=1
\documentclass[11pt, a4paper]{article}
\usepackage{amsmath, amssymb, amsthm}
\usepackage{geometry}
\usepackage{booktabs}
\usepackage{graphicx}
\usepackage{makecell}
\usepackage{multirow}
\usepackage{natbib}
\usepackage{float}
\usepackage{setspace}
\usepackage{textcomp}
\newtheorem{prop}{Proposition}
\newtheorem{proposition}[prop]{Proposition}
\newtheorem{assumption}{Assumption}

\title{\textbf{Estimating the housing production function \\with unobserved local heterogeneity}\thanks{I am grateful to Jiro Nemoto, Takanori Adachi, and Tatsuhito Kono for their insightful comments. This work is supported by the Japan Society for the Promotion of Science (JSPS) KAKENHI Grant Number 26K16373.}}
\author{Yusuke Adachi\thanks{College of Economics, Ritsumeikan University. E-mail: yadachi@fc.ristumei.ac.jp.}}
\date{\today}

\begin{document}
\maketitle

\begin{abstract}

Housing supply in dense cities depends on the ability of builders to substitute capital for scarce land. This margin is difficult to estimate because builders choose capital after observing microgeographic conditions that are only imperfectly observed by researchers. This paper develops a method for estimating revenue-based housing production functions in this setting. Because observed capital variation reflects both technological substitution and endogenous responses to latent local conditions, existing estimators can transmit unobserved heterogeneity into the estimated production function. The method treats the unobserved local conditions that affect capital choice as a scalar Markov state and combines the capital share equation with Markov moments implemented using repeated cross-sectional construction data. Monte Carlo simulations show that the estimator recovers capital and land elasticities under flexible production technologies when capital choices respond to latent local conditions. An application to newly constructed housing in Tokyo’s 23 special wards illustrates how the method can be implemented in a dense single-city setting. The results show that explicitly modeling latent local heterogeneity matters for estimated capital-land elasticities and implies returns to scale close to one.

\end{abstract}

\clearpage

\section{Introduction}

Housing supply in dense cities often expands not by using more land, but by using land more intensively. When developable land is scarce, builders add housing by substituting capital for land through larger, taller, or more capital-intensive structures. This capital-land substitution margin is therefore central to the price elasticity of new housing supply \citep{Saiz2010-qm, Baum-Snow2024-tb}, the vertical and density responses of urban development to changes in land values and demand \citep{Ahlfeldt2018-xd, Combes2019-sl}, and the welfare consequences of land-use regulations that restrict development intensity \citep{Bertaud2005-zd, Turner2014-rc, Hsieh2019-no, Koster2023-kx}. The housing production function is the structural object that summarizes this margin.

Estimating this object is difficult because builders choose capital after observing microgeographic conditions that are only imperfectly observed by the econometrician. These conditions include site amenities, neighborhood demand, local prices, regulatory constraints, and other features that affect expected development returns. The econometrician observes the resulting capital choice, but not the full set of conditions behind it. Observed capital choices therefore reflect both technological substitution between capital and land and builders' endogenous responses to latent local conditions. An estimator that treats this capital variation as purely technological attributes those responses to the production function. This transmission of unobserved conditions through capital choices is the bias emphasized by \citet{Ackerberg2015-tz} and \citet{Gandhi2020-kx}. Evidence from housing markets suggests that this problem is empirically relevant. For instance, productivity and related local conditions vary substantially across metropolitan areas \citep{Albouy2018-bm} and within metropolitan areas at fine spatial scales \citep{Baum-Snow2024-tb}. Because housing production functions are estimated from parcel-level development choices, heterogeneity at this scale is not merely a residual disturbance but a central identification problem.

This paper addresses this problem by explicitly controlling for the unobserved local conditions that enter builders' capital choices and by recovering both capital and land elasticities of the housing production function. The estimator separates the revenue-relevant production object from builders' endogenous responses to latent local conditions under a scalar-state restriction, using the capital share equation and Markov moments. The state restriction treats the unobserved local conditions that affect capital choice as a scalar Markov state. The capital share equation, implied by the builder's first-order condition, identifies the capital elasticity because the latent state drops out of the share equation. Integrating this elasticity with respect to capital recovers the production function up to a land-specific component, and Markov moments recover this remaining component and the state transition process. In repeated cross-sectional housing construction data, where the same parcel is rarely observed over time, matched previous-period observations provide the empirical lags used to implement these moments. This structure allows the estimator to recover both capital and land elasticities without external instruments for capital and without imposing constant returns to scale.

The proposed approach differs from existing housing production estimators in how it treats the latent local state behind observed development choices. Existing methods recognize that local productivity and parcel-level heterogeneity matter, but they do not impose an explicit dynamic structure on the state observed by builders. \citet{Epple2010-rg} exploit the duality between the production and cost functions and recover the production function from the equilibrium relationship between housing value per unit of land and unit land price. This approach is powerful under constant returns to scale, but the duality inversion no longer recovers the structural parameters once returns to scale are allowed to decrease. \citet{Combes2021-pu} relax the constant returns to scale restriction by integrating the first-order condition for capital. Their method accounts for location-level heterogeneous productivity and uses predicted inputs rather than external instruments to address unobserved parcel-level heterogeneity. The proposed approach differs by explicitly modeling the latent state that enters builders' capital choices.\footnote{In a within-metropolitan application of \citet{Combes2021-pu}, credible external instruments for capital are often difficult to obtain because the analysis relies primarily on variation within the same local housing market.} Our approach imposes a scalar-state restriction on local conditions and uses Markov moments to discipline the unobserved state.

The Monte Carlo simulations evaluate the finite-sample performance of the proposed estimator in environments where capital choices respond to latent local conditions. The exercises ask two questions. First, when capital choices respond to unobserved local conditions, do existing estimators recover the production technology or do they absorb part of this response? Second, can the proposed estimator recover both capital and land elasticities when the production technology is flexible? To answer these questions, we simulate data under Cobb--Douglas and constant elasticity of substitution (CES) technologies, each under constant and decreasing returns to scale, with a location-level latent state observed by builders. The Cobb--Douglas designs provide a benchmark in which the capital elasticity is constant. The CES designs are more demanding because elasticities vary across observations and capital choices respond to the latent state. The results show that both estimators perform well in the Cobb--Douglas benchmarks. In the CES designs, however, the estimator proposed by \citet{Combes2021-pu} exhibits transmission bias, whereas the proposed estimator recovers both capital and land elasticities with very small bias in sufficiently large samples. These results show that the estimator separates technological substitution from endogenous capital responses to unobserved local conditions in the settings where this distinction matters most.

The empirical application illustrates how the estimator can be implemented in an actual dense housing market. We apply the estimator to newly constructed housing in Tokyo's 23 special wards, where development occurs under microgeographic variation in amenities, regulation, demand, and land prices. The application combines administrative construction records, land-use survey data, and roadside land price data. Because housing value is not directly observed, the empirical implementation constructs a housing value from capital and land costs using the zero-profit condition. This construction follows the logic of the prior literature and provides a feasible zero-profit-implied revenue object for estimation. The paper therefore reports sensitivity exercises that clarify how deviations from the zero-profit-implied value affect the estimated elasticities. The results indicate that explicitly modeling latent local heterogeneity can matter quantitatively for the recovered production elasticities. The proposed estimator yields higher capital elasticities than those from \citet{Combes2021-pu}, especially in the land-intensive single-family segment. Because the estimator also recovers land elasticities, returns to scale can be assessed rather than imposed. The recovered elasticities are close to constant returns to scale, so constant returns emerge as an empirical implication under the maintained value construction rather than as a maintained assumption.

This paper contributes to two strands of literature. The first is the estimation of housing production functions. The classical monocentric-city tradition \citep{Muth-1969, Mills1967-dn} builds on a production technology combining land and capital, whose substitution properties govern equilibrium density and prices \citep{Rosen1974-uz}. Early empirical work focused on the elasticity of substitution between land and non-land inputs \citep{McDonald1981-fz, Jackson1984-au, Thorsnes1997-vc}. Recent work develops structural methods that recover the production function from observable equilibrium data. \citet{Epple2010-rg} exploit duality under constant returns to scale, and \citet{Combes2021-pu} relax this restriction by integrating the capital first-order condition. Complementary work has pushed in related directions, including long-run estimation using tall buildings \citep{Ahlfeldt2018-xd}, neighborhood-level supply elasticities with an intensive-extensive margin decomposition \citep{Baum-Snow2024-tb}, and metropolitan-level housing productivity \citep{Albouy2018-bm}. Our contribution is to formalize the role of unobserved location-level heterogeneity in this framework and provide an estimator designed for settings in which flexible technologies make capital choices informative about the latent local state.

The second strand is the welfare analysis of land-use regulation. A theoretical tradition in the monocentric model shows that the welfare cost of floor-area-ratio and height restrictions depends on how binding they are on the capital-intensive margin \citep{Brueckner2012-km}. Empirical work has quantified the wedge between housing prices and construction costs generated by regulation in dense urban areas \citep{Glaeser2005-cy,Brueckner_2017,Adachi_2024}, and more recent structural analyses evaluate the welfare effects of specific zoning reforms \citep{Anagoletal2021, Kulka2026-ez}. At the aggregate level, recent spatial-equilibrium analyses quantify the macroeconomic consequences of housing supply constraints \citep{Herkenhoff2018-nd, Parkhomenko2023-yp}, while other work examines how supply responds dynamically to demand shocks \citep{Murphy2018-xl}. The welfare cost of these regulations is tied to the output elasticity of capital in the housing production function. Holding other primitives fixed, restrictions on capital-intensive development are more costly when the capital margin is more productive.

The remainder of the paper is organized as follows. Section~\ref{sec:model} presents the model and the two-step estimation strategy. Section~\ref{sec:monte_carlo} reports Monte Carlo evidence on finite-sample performance. Section~\ref{sec:empirical} describes the data construction and the empirical results. Section~\ref{sec:conclusion} concludes.

\section{Model and Estimation Strategy}\label{sec:model}

\subsection{The Model}
The model formalizes the timing behind the identification problem introduced above. Builders choose capital after observing local conditions that affect expected development returns, while the econometrician observes the resulting input choices but not all of the conditions behind them. The purpose of the model is to show how this timing makes capital endogenous to a latent local state and how the share equation and state restriction can be used to control for it.

We assume that the relationship between output and inputs is determined by an underlying production function and a latent local state. Builders operate in a competitive housing market and decide how much capital to use on parcels whose size is predetermined for the capital choice.\footnote{This model builds on the housing production models developed by \citet{Epple2010-rg} and \citet{Combes2021-pu}.}
To simplify notation, we index each observed construction by a location-time cell $(j,t)$ and suppress a separate construction-level index.\footnote{This notation can be interpreted as defining locations at a sufficiently fine spatial scale that an observed construction represents the local development environment in a location-period. It does not require that every neighborhood-period literally contains only one observed construction; rather, the empirical implementation approximates the relevant local state using nearby observations.} A builder in location $j$ at time $t$ produces housing services $Y_{jt}$ from two inputs: parcel size $L_{jt}$, which is predetermined for the capital choice, and capital $K_{jt}$, which summarizes all non-land inputs such as construction materials, structural components, and labor.\footnote{The model conditions on the developed parcel and does not model the extensive margin of parcel selection or land assembly.} The price of capital, $P_k$, is constant across locations within a metropolitan area and normalized to one. The key implication is that land is the predetermined input, whereas capital is the builder's endogenous adjustment margin. The latent local state in the model should be interpreted more broadly than pure physical productivity. Physical productivity enters the production technology through a location- and time-specific component, denoted $\omega_{jt}$. Local housing prices also affect the revenue product of capital. Other local conditions, such as site amenities, neighborhood demand, and the regulatory environment, can also affect the profitability of deploying capital on a parcel. Our treatment of location-level heterogeneity is in the spirit of \citet{Baum-Snow2024-tb}, who introduce productivity into their model of housing supply, but the estimation below controls for the composite local state relevant for revenue and capital choice rather than separately identifying each primitive component.

The production function is specified so that the substitution technology between capital and land is separated from local productivity. Housing services are produced according to
\begin{equation*}
Y_{jt} = H(K_{jt}, L_{jt}) \exp(\omega_{jt} + \varepsilon_{jt}).
\end{equation*}
The function $H(\cdot)$ combines capital and land. The component $\omega_{jt}$ is a location-time productivity term that is common to builders operating in location $j$ at time $t$, and it enters production in a Hicks-neutral way. The shock $\varepsilon_{jt}$ is idiosyncratic and is realized after the capital choice. We do not impose a functional form on $H(\cdot)$, allowing the production technology to exhibit flexible substitution between capital and land and general returns to scale.\footnote{Specifically, we allow for constant or decreasing returns to scale, consistent with standard assumptions in the housing literature. In addition, increasing returns to scale would be inconsistent with competitive pricing.} This specification keeps the technological substitution margin inside $H(\cdot)$, while allowing local productivity to shift the expected return to capital and hence the builder's capital choice.

The empirical state used in the estimation is revenue based because housing services and housing prices are not separately observed. The data do not provide physical housing services $Y_{jt}$ separately from their price $P^h_{jt}$. The object linked to input choices is housing value,
\begin{equation*}
V_{jt} \equiv P^h_{jt}Y_{jt}.
\end{equation*}
For this reason, physical productivity cannot be separately recovered from the local component of housing prices using housing value data alone. 
With the average price component absorbed into the intercept of the revenue-based production object, write log housing prices as
\begin{equation*}
\log P^h_{jt}=p^h_{jt}+\nu_{jt}.
\end{equation*}
The term $p^h_{jt}$ is a location-time component, and $\nu_{jt}$ is an idiosyncratic price shock.\footnote{Equivalently, if one writes $\log P^h_{jt}=\mu_p+p^h_{jt}+\nu_{jt}$, the average price component $\mu_p$ is absorbed into the intercept of the revenue-based production object, $p^h_{jt}$ is included in the composite state $\tilde{\omega}_{jt}$, and $\nu_{jt}$ is collected into the ex post revenue shock.} With the average price component absorbed into the intercept of the revenue-based production object and the idiosyncratic component treated as part of the ex post revenue shock, the local state relevant for expected revenue and capital choice is
\begin{equation*}
\tilde{\omega}_{jt} \equiv \omega_{jt} + p^h_{jt}.
\end{equation*}
This composite state is not pure physical productivity. It is the revenue-relevant local condition that builders observe when choosing capital.\footnote{This composite state is analogous to revenue productivity in the production-function literature \citep{DELOECKER2021141}. It combines physical productivity with the price component that cannot be separately identified from revenue data alone.} Local amenities, neighborhood demand, the regulatory environment, and other microgeographic factors may affect this state through productivity, through prices, or through both. The estimation below controls for $\tilde{\omega}_{jt}$ rather than separately identifying each primitive component. The Markov restriction introduced below is imposed on this composite state, not on physical productivity or local prices separately. For notational simplicity, we collect the idiosyncratic production shock and the idiosyncratic price shock into a single ex post revenue shock, $\epsilon_{jt} \equiv \varepsilon_{jt}+\nu_{jt}$. This shock is realized after the capital choice and is normalized so that $\mathbb{E}[\exp(\epsilon_{jt})\mid I_{jt}]=1$. With this normalization, the revenue equation can be written as
\begin{equation*}
V_{jt} = H(K_{jt},L_{jt}) \exp(\tilde{\omega}_{jt} + \epsilon_{jt}).
\end{equation*}

Given the revenue-based state, the builder's problem links the unobserved local condition to the observed capital choice. When choosing capital, the builder observes parcel size $L_{jt}$ and the composite state $\tilde{\omega}_{jt}$, but not the ex post shock. Let $I_{jt}$ denote the information available at the time of the capital decision. The builder chooses capital to maximize expected profit,
\begin{equation*}
\max_{K_{jt}} E \left[ H(K_{jt},L_{jt}) \exp(\tilde{\omega}_{jt}+\epsilon_{jt}) - K_{jt} - R_{jt} \mid I_{jt} \right],
\end{equation*}
where $R_{jt}$ is the land acquisition cost and the price of capital has been normalized to one. The first-order condition for an interior capital choice is\footnote{The maintained first-order condition applies to the intensive capital margin under the prevailing regulatory environment. If development constraints bind directly on capital or floor area, the observed choice may include a regulatory wedge. The empirical analysis below interprets the recovered object as a revenue-based production object under the prevailing regulatory regime.}
\begin{equation}
\frac{\partial H(K_{jt},L_{jt})}{\partial K_{jt}} \exp(\tilde{\omega}_{jt}) = 1. \label{eq:foc_capital}
\end{equation}
Equation~\eqref{eq:foc_capital} shows why capital is endogenous to the latent local state. Holding parcel size fixed, a higher value of $\tilde{\omega}_{jt}$ raises the expected revenue product of capital and changes the builder's optimal capital input. Thus, observed capital choices are functions of both predetermined land and the unobserved local state,
\begin{equation*}
K_{jt} = K_t(L_{jt},\tilde{\omega}_{jt}).
\end{equation*}
This relationship is the source of the identification problem. The same capital variation used to estimate $H(\cdot)$ also reflects builders' responses to $\tilde{\omega}_{jt}$. To control for this response, we impose structure on the latent local state.

\begin{assumption}[Markov process]\label{as:markov}
The payoff-relevant composite state $\tilde{\omega}_{jt}$ follows a first-order Markov process:
\begin{equation*}
    \tilde{\omega}_{jt} = g(\tilde{\omega}_{j,t-1}) + \eta_{jt},
\end{equation*}
where $\eta_{jt}$ is an innovation satisfying $\mathbb{E}[\eta_{jt} \mid I_{j,t-1}] = 0$, and $I_{j,t-1}$ denotes the information common to builders in location $j$ available at the end of period $t-1$.
\end{assumption}

Assumption~\ref{as:markov} is imposed directly on the composite state relevant for capital demand, not separately on physical productivity and housing prices. It does not require each primitive determinant of profitability to follow its own Markov process, nor does it assume that the sum of two independently specified Markov processes is automatically Markov. Rather, the assumption is that the one-dimensional state $\tilde{\omega}_{j,t-1}$ is a sufficient statistic for forecasting the payoff-relevant component $\tilde{\omega}_{jt}$ given the information set $I_{j,t-1}$. This restriction is natural in a housing setting because the main determinants of local development returns are spatially and temporally persistent. Site characteristics and neighborhood amenities are durable, the regulatory environment changes infrequently, and local demand and price conditions typically evolve gradually. The assumption is most plausible when the sample period does not contain large unanticipated shocks, such as major disasters or abrupt regulatory regime changes, that would break the link between current and lagged local profitability. The innovation $\eta_{jt}$ captures changes in local profitability that are not predictable from the lagged state.

The Markov restriction describes how the latent state evolves over time. We also need a restriction that links this state to observed input choices.

\begin{assumption}[Scalar state and monotonicity]\label{as:scalar}
Conditional on parcel size, the local conditions relevant for the builder's capital choice can be summarized by a scalar unobservable state $\tilde{\omega}_{jt}$. Capital demand is strictly increasing in this state for all $L_{jt}$ in the support:
\begin{equation*}
    K_{jt} = K_t(L_{jt}, \tilde{\omega}_{jt}).
\end{equation*}
\end{assumption}

Assumption~\ref{as:scalar} is the step that makes observed capital informative about the latent state. It has two parts. The first is a scalar-state restriction. This restriction does not require site productivity, amenities, neighborhood demand, local price conditions, and the regulatory environment to share a common primitive source. It requires only that their combined effect on the builder's capital choice can be represented by one state variable. This is the sense in which the model treats unobserved local heterogeneity as scalar.

The second part is monotonicity. The housing setting gives this restriction a natural interpretation. Parcel size $L_{jt}$ is predetermined, while capital $K_{jt}$ is the builder's adjustable input. Holding parcel size fixed, a higher value of $\tilde{\omega}_{jt}$ raises the expected return to deploying capital on the parcel. Under an interior optimum and strict concavity in capital, the builder responds by choosing more capital. Capital demand is therefore strictly increasing in the latent state. This monotone relationship implies that, conditional on land, the observed capital choice can be inverted to recover the latent state:
\begin{equation*}
    \tilde{\omega}_{jt} = \psi_t(K_{jt}, L_{jt}).
\end{equation*}
The inversion is the key implication of Assumption~\ref{as:scalar}. It allows the latent local state to be controlled for through observed input choices, following the logic of \citet{Levinsohn2003-td} and \citet{Ackerberg2015-tz}, but adapted to a housing setting in which land is parcel-specific and capital is the relevant adjustment margin.

The model has a direct implication: observed capital choices cannot be interpreted as pure variation in capital-land substitution. The composite state $\tilde{\omega}_{jt}$ affects the expected return to capital, so builders in more favorable local conditions choose different capital inputs even for parcels with the same land input. Under free entry, the same state is also capitalized into land acquisition costs. The variables used to estimate the production function therefore carry information about both the substitution technology and unobserved local conditions. This is the source of transmission bias in the present setting. If the econometrician does not control for $\tilde{\omega}_{jt}$, the response of capital to unobserved local conditions is transmitted into the estimated production function. An estimator may then attribute capital differences to technological substitution, even when they partly reflect builders' responses to unobserved local conditions. Appendix B shows this mechanism formally for the estimators of \citet{Epple2010-rg} and \citet{Combes2021-pu}. The estimation strategy developed next addresses this bias by controlling for the composite state while using the share equation and the Markov restriction to identify the production function.

\subsection{Estimation Strategy}
The key idea is to identify the part of the production function that does not require conditioning on the latent state, and then use the state restriction to recover the remaining part. The estimation strategy implements this idea in two steps. The first step uses the capital share equation implied by the builder's first-order condition. In this equation, the latent composite state drops out, so the capital elasticity can be identified from observed inputs and the capital share. The second step uses the Markov restriction to recover the remaining components of the production function. Integrating the estimated capital elasticity recovers the production function up to a land-specific integration component. This remaining component is then identified jointly with the Markov transition function using moment conditions based on predetermined land input.

\subsubsection*{Step 1: Recovering the output elasticity with respect to capital}
Step 1 identifies the output elasticity with respect to capital from the capital share equation implied by the builder's first-order condition. Let $V_{jt} \equiv P^h_{jt}Y_{jt}$ denote housing value, and define the observed capital share in housing value as
\begin{equation*}
    s_{jt} \equiv \frac{K_{jt}}{V_{jt}}.
\end{equation*}
We use the logarithmic notation $k_{jt} \equiv \log K_{jt}$ and $l_{jt} \equiv \log L_{jt}$, and write $h(k,l)$ for the log production function. The output elasticity with respect to capital is
\begin{equation*}
    D(k,l) \equiv \frac{\partial h(k,l)}{\partial k}.
\end{equation*}
The first-order condition links this elasticity to the capital share. Multiplying the first-order condition by $K_{jt}$ and dividing by expected housing value gives the capital elasticity. Because observed housing value contains the ex post shock $\epsilon_{jt}$, the observed share satisfies
\begin{equation*}
    \log s_{jt} = \log D(k_{jt},l_{jt}) - \epsilon_{jt}.
\end{equation*}
The composite state $\tilde{\omega}_{jt}$ drops out of this equation. Because $\epsilon_{jt}$ is realized after input choices, it is orthogonal to the observed inputs used in the share equation. We approximate $D(k,l)$ by a truncated polynomial series and estimate the share equation by nonlinear least squares. This first step delivers an estimate of the elasticity function, $\hat{D}(\cdot)$, together with the implied residual $\hat{\epsilon}_{jt}$.

\subsubsection*{Step 2: Estimating the remaining components}
Step 2 recovers the part of the production function that is not identified by the share equation. Step 1 identifies the derivative of the log production function with respect to log capital, $D(k,l)=\partial h(k,l)/\partial k$. Therefore, there exists a land-specific component $C(l)$ such that
\begin{equation*}
    h(k,l) = \int^k D(\kappa,l)\,d\kappa + C(l),
\end{equation*}
where the integral is taken with respect to log capital, holding land fixed. The integration function $C(l)$ includes any component of the revenue-based production function that is not identified from capital variation, including its intercept. Hence, $C(l)$ can be written as the sum of a constant and a land-specific component: $C(l)=c+C_0(l)$, where $c$ captures the level of the revenue and $C_0(l)$ captures the part that varies with land. Once $C(l)$ is recovered, the land elasticity is obtained as
\begin{equation*}
D_l(k,l) = \frac{\partial h(k,l)}{\partial l} = \frac{\partial}{\partial l} \left[ \int^k D(\kappa,l)\,d\kappa \right] + C'(l).
\end{equation*}
An additive constant in $C(l)$ therefore does not affect the land elasticity or returns to scale.

Let $v_{jt}\equiv \log V_{jt}$ denote log housing value. Using the estimated elasticity from Step 1, define the adjusted log housing value as
\begin{equation*}
    \hat{\phi}_{jt} \equiv v_{jt} - \hat{\epsilon}_{jt} - \int^{k_{jt}} \hat{D}(\kappa,l_{jt})\,d\kappa,
\end{equation*}
where $\hat{\epsilon}_{jt}$ is the implied ex post shock from the share equation, defined with the sign convention $\hat{\epsilon}_{jt}=\log \hat{D}(k_{jt},l_{jt})-\log s_{jt}$. After the capital component and the ex post revenue shock have been removed, the adjusted value can be written as
\begin{equation*}
    \hat{\phi}_{jt} = C(l_{jt}) + \tilde{\omega}_{jt} + \xi_{jt},
\end{equation*}
where $\xi_{jt}$ collects first-step estimation error and series approximation error. The level component of the revenue-based object is included in $C(l)$.

A cross section alone cannot separate these two objects. A high value of $\hat{\phi}_{jt}$ may reflect either a larger land-specific component $C(l_{jt})$ or a more favorable local state $\tilde{\omega}_{jt}$. The Markov restriction provides the additional structure needed for this separation. Given a candidate function $C(\cdot)$, each observation provides a proxy for the location-time state:
\begin{equation*}
    \hat{\omega}_{jt}(C) \equiv \hat{\phi}_{jt} - C(l_{jt}).
\end{equation*}
The Markov restriction itself applies to the true location-time state:
\begin{equation*}
    \tilde{\omega}_{jt} = g(\tilde{\omega}_{j,t-1}) + \eta_{jt}, \quad \mathbb{E}[\eta_{jt} \mid I_{j,t-1}] = 0.
\end{equation*}
Because the true state and its lag are unobserved, the empirical implementation uses proxy states constructed from observed constructions. In repeated cross-sections, the same parcel is not observed over time, so the lagged state $\tilde{\omega}_{j,t-1}$ is approximated using a nearby observation from the previous period. The proxy-based Markov residual is
\begin{equation*}
    \hat{\eta}_{jt}(C,g) = \hat{\omega}_{jt}(C) - g\left(\hat{\omega}_{m(j,t)}(C)\right),
\end{equation*}
where $\hat{\omega}_{m(j,t)}(C)$ is the matched proxy for the lagged local state. Using the matched previous-period observation as the empirical lag, the predetermined nature of lagged land input gives moment conditions of the form
\begin{equation*}
    \mathbb{E}\left[\hat{\eta}_{jt}(C,g) z_{m(j,t)}\right]=0,
\end{equation*}
where $z_{m(j,t)}$ is a vector of functions of lagged land input from the matched previous-period observation. These moments use variables measured before the current innovation $\eta_{jt}$. In the empirical quasi-panel, $z_{m(j,t)}$ is not the lagged input of the same parcel but the lagged input of a nearby previous-period construction. The validity of this implementation therefore relies on the local-homogeneity approximation that nearby constructions serve as proxies for the same underlying location-level state.

The Markov restriction, combined with the share-equation residualization, provides the additional structure needed to separate the land-specific component from the composite local state. Step 2 estimates $C(\cdot)$ and $g(\cdot)$ jointly by choosing them so that the Markov moments are close to zero. In the empirical implementation, both functions are approximated by low-order polynomials, as detailed in Appendix A.1. This identification uses the share equation and the Markov restriction; the matched-lag construction is the practical device that implements the Markov moments when the same parcel is not observed repeatedly.

Two data implementation issues remain after the identification argument. The first issue is how to construct lagged state proxies when parcel panels are unavailable. Housing construction data are usually repeated cross-sections and do not track the same parcel over time. When precise building coordinates are available, we implement the matched-lag construction by matching each observation in period $t$ to its geographically nearest observation in period $t-1$ using building centroid coordinates.\footnote{When such local matching is unavailable, the same logic can be implemented with location-time cell averages, as in the pseudo-panel construction discussed in Appendix~\ref{app:mc_quasi_pseudo}.} This construction relies on local homogeneity. Within sufficiently small neighborhoods, nearby observations provide information about the same underlying local state. The second issue is that the variables required by the estimator are not always observed jointly. In the empirical application, housing value is not directly observed. We therefore construct the value measure used in estimation from capital and land costs using the zero-profit accounting relationship,
\begin{equation*}
V^{zp}_{jt}=K_{jt}+R_{jt}.
\end{equation*}
This variable is a zero-profit-implied revenue measure rather than an observed transaction value.\footnote{The estimator is applied to this constructed revenue in the empirical application below, and the empirical analysis reports sensitivity exercises that examine how the estimated elasticities change when the constructed value deviates from the maintained zero-profit-implied measure.}

\section{Monte Carlo Simulation}\label{sec:monte_carlo}
\subsection{Design}

The Monte Carlo exercises evaluate the finite-sample performance of the proposed estimator in environments where capital choices respond to latent local conditions. The simulations ask whether the estimator recovers the relevant input elasticities when the maintained scalar-state and Markov restrictions hold, and whether the estimator suggested by \citet{Combes2021-pu} absorb part of builders' endogenous responses to the latent state.

Specifically, we simulate four designs: Cobb--Douglas and CES technologies, each under constant and decreasing returns to scale. The Cobb--Douglas designs provide a benchmark in which the capital elasticity is constant, so both estimators should perform well. The CES designs are more demanding because elasticities vary with observed inputs, making it easier for builders' responses to the latent state to be attributed to the production function. We consider both constant returns to scale (CRS) and decreasing returns to scale (DRS) in each technology, and simulate each design with 500 replications. Each replication generates data at the location-time level, with $J = 1{,}000$ locations, one construction per location in each period, and $T = 5$ periods. This design matches the notation in the model, where each observed construction is treated as a location-time observation. The cross-sectional dimension $J$ provides variation across locations needed to identify the location-level latent state, while the time-series dimension $T$ provides the within-location variation needed to recover the law of motion for the latent state. To construct lagged location-level states in the simulations, each period-$t$ observation is linked to the observation from the same location in period $t-1$. This is the idealized counterpart of the nearest-neighbor quasi-panel used in the empirical application, where the same physical parcel is generally not observed repeatedly and lagged local states are proxied by nearby previous-period observations.

The simulated location-level latent state follows a first-order Markov process:
\begin{equation*}
    \omega_{jt} = \delta_0 + \delta_1 \omega_{j,t-1} + \eta_{jt},
\end{equation*}
with $(\delta_0, \delta_1) = (0.5, 0.5)$, $\eta_{jt} \sim N(0, 0.1^2)$, and initial state $\omega_{j1} \sim U[0.8, 1.2]$.
This construction is consistent with Assumption \ref{as:markov}. In the simulations, the latent state can be interpreted as the composite revenue-productivity state used in the estimation.

\subsubsection*{Cobb--Douglas production function}
Output under Cobb--Douglas technology is generated according to
\begin{equation*}
    Y_{jt} = K_{jt}^{\beta_k} L_{jt}^{\beta_l} \exp(\omega_{jt} + \epsilon_{jt}).
\end{equation*}
Builders choose capital to maximize expected profit taking $P^h_{jt}$, $L_{jt}$, and $\omega_{jt}$ as given, with $\epsilon_{jt}$ realized after the input choice. The resulting capital demand is
\begin{equation*}
    K_{jt} = \left( P^h_{jt} \beta_k L_{jt}^{\beta_l} \exp(\omega_{jt}) \right)^{\frac{1}{1-\beta_k}}.
\end{equation*}
Under CRS, we set $(\beta_k, \beta_l) = (0.6, 0.4)$; under DRS with returns-to-scale parameter $\nu = \beta_k + \beta_l = 0.85$, we set $(\beta_k, \beta_l) = (0.51, 0.34)$.

\subsubsection*{CES production function}
Output under CES technology is generated according to
\begin{equation*}
    Y_{jt} = \left( \beta_k K_{jt}^{\rho} + \beta_l L_{jt}^{\rho} \right)^{\nu/\rho} \exp(\omega_{jt} + \epsilon_{jt}),
\end{equation*}
where $\rho$ governs the elasticity of substitution $\sigma = 1/(1-\rho)$ and $\nu$ controls returns to scale. We set $\rho = 0.5$ so that $\sigma = 2$, and $(\beta_k, \beta_l) = (0.4, 0.6)$ in both CRS and DRS designs. Under CRS, $\nu = 1$; under DRS, $\nu = 0.85$. As in the Cobb--Douglas case, builders choose capital to maximize expected profit, with $\epsilon_{jt}$ realized after the input choice. Under CES, the output elasticity with respect to capital varies across observations. When evaluating estimator performance, we therefore compare the estimates against the sample-average capital elasticity implied by the simulated data, rather than against a single common parameter. This makes the evaluation comparable across Cobb--Douglas and CES environments while preserving the underlying heterogeneity in elasticities.

\subsection{Results}

Table \ref{tab:mc_results} reports the Monte Carlo results for the estimated output elasticities. The table is organized to distinguish the target object from the estimator's sampling behavior. The column ``Target'' reports the target elasticity used for evaluation. Under Cobb--Douglas, this target is a single parameter common to all observations and replications. Under CES, true elasticities vary across observations, so the target in each replication is the sample average of the observation-level true elasticities. The table reports the average of this target across replications, and the bracketed range gives its 2.5th and 97.5th percentiles across replications. The column ``Mean'' reports the average estimated elasticity across Monte Carlo replications. The column ``Bias'' is the difference between ``Mean'' and ``Target,'' using the corresponding simulation averages. The column ``SD'' reports the simulation standard deviation of the point estimates across replications. Finally, ``95\% Cov'' reports the Monte Carlo coverage rate, defined as the fraction of replications in which the target elasticity lies within a simulation-based normal interval centered at the estimate.

Table~\ref{tab:mc_results} reports the baseline Monte Carlo results. The proposed estimator recovers both input elasticities accurately across the Cobb--Douglas and CES designs. In the Cobb--Douglas benchmarks, both the proposed estimator and the CDG estimator recover the capital elasticity well. In the CES designs, where elasticities vary across observations and capital choices respond to the latent state, the proposed estimator continues to recover the average capital and land elasticities, whereas the estimator suggested by \citet{Combes2021-pu} (hereafter CDG) exhibits transmission bias in the capital elasticity. The recovered returns to scale are also close to the true values; in the decreasing-returns designs, the largest returns-to-scale bias is below 0.003.

The first result is that both estimators perform well in the Cobb--Douglas designs, and the proposed estimator is at least as accurate as CDG for the capital elasticity. Under CD (CRS), the true capital elasticity is 0.600. CDG yields a mean of 0.602, while the proposed estimator recovers a mean of 0.600. The proposed estimator also accurately recovers the land elasticity: the true value is 0.400, the mean estimate is 0.398, the bias is $-0.002$, and the Monte Carlo coverage rate is 0.95. Under CD (DRS), the same pattern continues. For capital, the true elasticity is 0.510, and both CDG and the proposed estimator recover a mean of 0.510. For land, the true elasticity is 0.340, the proposed mean is 0.337, the bias is $-0.003$, and the Monte Carlo coverage rate is 0.94. Taken together, these results show that in the Cobb--Douglas cases both approaches recover the capital elasticity well, and the proposed estimator additionally recovers the land elasticity with small bias.

The second result is that the proposed estimator continues to perform well under CES, whereas CDG has low Monte Carlo coverage rates for the CES target elasticities. Under CES (CRS), the true average capital elasticity is 0.442, with a 95\% range of [0.437, 0.447]. The proposed estimator recovers this target almost exactly: the mean estimate is 0.442, the bias is 0.000, the standard deviation is 0.003, and the Monte Carlo coverage rate is 0.96. CDG, by contrast, yields a mean estimate of 0.478, implying a bias of 0.036, and its Monte Carlo coverage rate falls to 0.00. The same pattern appears in the Monte Carlo coverage comparison under CES (DRS). The true average capital elasticity is 0.181, with a 95\% range of [0.179, 0.183]. The proposed estimator again places the target inside the simulation-based interval in almost all replications, with mean 0.181, bias 0.000, standard deviation 0.001, and Monte Carlo coverage 0.95. CDG has a small average bias in this design, with mean 0.184 and bias 0.003, but its Monte Carlo coverage rate is only 0.24. The land elasticity estimates from the proposed estimator also have high Monte Carlo coverage rates in both CES designs: under CES (CRS), the true value is 0.558 and the mean estimate is 0.559, with bias 0.001 and Monte Carlo coverage 0.95; under CES (DRS), the true value is 0.669 and the mean estimate is 0.671, with bias 0.003 and Monte Carlo coverage 0.96. The low Monte Carlo coverage rates for CDG in the CES designs reflect transmission bias in the flexible-technology setting. When capital choices respond to an unobserved composite state, integrating the first-order condition without explicitly controlling for that state can cause the recovered production function to inherit unobserved heterogeneity. This source of bias is not apparent in the Cobb--Douglas benchmark but matters under CES, where the relevant criterion is whether the estimator reliably covers the target elasticity across replications rather than whether the simulation average happens to be close in a particular design.

Additional simulations are reported in Appendix~\ref{app:mc_additional}. These simulations, for instance, examine spatial correlation in the latent state, deviations from the zero-profit-implied value object, and violations of the maintained scalar-state and Markov restrictions. These exercises are sensitivity checks rather than tests of the maintained assumptions. They show that random or location-level value deviations have limited effects on the estimates, while deviations correlated with land prices or capital intensity can distort the land elasticity and returns to scale. In addition, the estimator of \citet{Epple2010-rg} (hereafter EGS) is omitted from the main table because it estimates a cost-function elasticity rather than a production-function elasticity. The two coincide only under CRS, so outside that case EGS is not directly comparable to the true object used here.

\begin{table}[H]
\centering
\caption{Monte Carlo Simulation Results}
\label{tab:mc_results}
\small
\begin{tabular}{lcccccc}
\toprule
 & Elasticity & Target & Mean & Bias & SD & 95\% Cov \\
\hline
\multicolumn{7}{l}{\textit{Panel A: Cobb--Douglas (CRS)}} \\
CDG      & $\widehat{\bar{D}}_k$ & 0.600 & 0.602 & 0.002 & 0.001 & 0.73 \\
Proposed & $\widehat{\bar{D}}_k$ & 0.600 & 0.600 & 0.000 & 0.001 & 0.95 \\
 & $\widehat{\bar{D}}_l$   & 0.400 & 0.398 & $-$0.002 & 0.032 & 0.95 \\[6pt]
\multicolumn{7}{l}{\textit{Panel B: Cobb--Douglas (DRS)}} \\
CDG      & $\widehat{\bar{D}}_k$ & 0.510 & 0.510 & 0.000 & 0.001 & 0.94 \\
Proposed & $\widehat{\bar{D}}_k$ & 0.510 & 0.510 & 0.000 & 0.001 & 0.95 \\
 & $\widehat{\bar{D}}_l$    & 0.340 & 0.337 & $-$0.003 & 0.032 & 0.94 \\[6pt]
\multicolumn{7}{l}{\textit{Panel C: Constant Elasticity of Substitution (CRS)}} \\
CDG      & $\widehat{\bar{D}}_k$ & \makecell{0.442 \\ {\small $[0.437,\,0.447]$}} & 0.478 & 0.036 & 0.004 & 0.00 \\[2pt]
Proposed & $\widehat{\bar{D}}_k$ & \makecell{0.442 \\ {\small $[0.437,\,0.447]$}} & 0.442 & 0.000 & 0.003 & 0.96 \\[2pt]
 & $\widehat{\bar{D}}_l$   & \makecell{0.558 \\ {\small $[0.553,\,0.563]$}} & 0.559 & 0.001 & 0.032 & 0.95 \\[6pt]
\multicolumn{7}{l}{\textit{Panel D: Constant Elasticity of Substitution (DRS)}} \\
CDG      & $\widehat{\bar{D}}_k$ & \makecell{0.181 \\ {\small $[0.179,\,0.183]$}} & 0.184 & 0.003 & 0.001 & 0.24 \\[2pt]
Proposed & $\widehat{\bar{D}}_k$ & \makecell{0.181 \\ {\small $[0.179,\,0.183]$}} & 0.181 & 0.000 & 0.001 & 0.95 \\[2pt]
 & $\widehat{\bar{D}}_l$    & \makecell{0.669 \\ {\small $[0.667,\,0.671]$}} & 0.671 & 0.003 & 0.031 & 0.96 \\
\bottomrule
\end{tabular}
\begin{minipage}{0.9\textwidth}\scriptsize
    \textit{Notes:} 500 Monte Carlo replications. one observed construction per location-period, $J = 1{,}000$ locations, $T = 5$ periods. No spatial correlation in the latent state. ``Proposed'' = GNR-based estimator; ``CDG'' = integration-based estimator of \citet{Combes2021-pu}. $\widehat{\bar{D}}_k$ and $\widehat{\bar{D}}_l$ denote estimated sample-average output elasticities. For Cobb--Douglas designs, ``Target'' is the common true elasticity. For CES designs, the target in each replication is the sample average of observation-level true elasticities; the number shown above the brackets reports the average target across replications, and brackets report its 2.5th and 97.5th percentiles across replications. ``Mean'' is the average estimated elasticity across replications, ``Bias'' is Mean minus Target, and ``SD'' is the simulation standard deviation of the estimates. ``95\% Cov'' is the Monte Carlo coverage rate, computed as the fraction of replications in which the target elasticity lies within $\hat{\theta}_r \pm 1.96 \times SD(\hat{\theta})$.
\end{minipage}
\vspace{0.1cm}
\end{table}

\section{Empirical Application}\label{sec:empirical}
The empirical application has two purposes: to implement the estimator in a realistic data environment and to examine whether controlling for unobserved local heterogeneity changes the estimated production elasticities. We study newly constructed housing in Tokyo's 23 special wards, which form the dense urban core of the metropolitan area. The empirical challenge is that the variables required for estimation are not jointly observed in a single dataset. We therefore combine multiple administrative sources to construct the key inputs and use the zero-profit condition to construct housing value when it is not directly observed.

\subsection{Data}
The data construction follows directly from the objects required by the estimator. We need capital input, lot size, land acquisition cost, and housing value, but these variables are not jointly available in a single Japanese dataset. We therefore combine three administrative sources: the Survey on Building Construction Started (\textit{Kenchiku Chakkou Toukei}), the Survey on Current Land Use (\textit{Tochi Riyo Genkyo Chosa}), and the Roadside Land Price survey (\textit{Rosenka}).

The first, the Survey on Building Construction Started, is the primary source of information on capital input. It is an administrative dataset compiled by the Ministry of Land, Infrastructure, Transport and Tourism (MLIT) from filings submitted when new buildings are started. For our analysis period (2012--2016), we use the data from each year, with the recorded year corresponding to construction completion. The survey reports construction cost, which we use as the measure of capital input, together with building attributes such as ward and small area code, total floor area, lot size, number of stories, and structure type. Its main limitation for our purposes is that it does not contain information on land acquisition cost. The second, the Survey on Current Land Use, provides the parcel-level continuity and location information needed to identify newly developed sites. It is a two-period panel for 2011 and 2016 that records detailed building characteristics together with precise geographic location. Its key advantage for our purposes is that repeated observations on the same parcel allow us to determine whether new construction occurred between the two survey waves. The third, the Roadside Land Price survey, provides the unit land values used to construct an approximation to parcel-level land acquisition cost. These roadside values (\textit{rosenka}) are administrative land assessment values published every three years by the National Tax Agency for inheritance and gift tax purposes, with each value defined as the assessed price per square meter of a standard parcel facing a given road. We use the 2012 and 2015 waves for our analysis.

We construct the estimation sample starting from newly developed parcels in the Survey on Current Land Use, to which we attach land prices and construction records from the other two sources. First, we assign each newly developed parcel in the Survey on Current Land Use a unit land value by matching to the nearest observation in the Roadside Land Price survey in both space and year. Since the Roadside Land Price survey is available only for 2012 and 2015, observations from 2012 and 2013 are matched to the 2012 wave, and those from 2014 through 2016 are matched to the 2015 wave. Denoting the matched roadside land price by $p^l_{jt}$, we approximate parcel-level land acquisition cost as
\begin{equation*}
R_{jt}=L_{jt}p^l_{jt}.
\end{equation*}
Second, we match these parcels to construction records in the Survey on Building Construction Started using nearest-neighbor matching, requiring exact agreement on ward code, number of stories, housing type, and structure type, and selecting the closest match on the basis of total floor area and lot size. This step supplies construction cost, which we use as the measure of capital input $K_{jt}$. 
Third, because housing value is not directly observed, we construct the value measure used in estimation from capital and land costs:
\begin{equation*}
V^{zp}_{jt}=K_{jt}+R_{jt}.
\end{equation*}
This variable is a zero-profit-implied value object rather than an observed transaction value. The estimator is therefore applied to a revenue object constructed from the equilibrium accounting relationship. The robustness exercises below and in the appendix examine how sensitive the estimated elasticities are to deviations from this constructed value measure. In the empirical application, $V_{jt}$ denotes this constructed value measure unless otherwise stated.

To ensure reliable matches, we impose quality thresholds of 20\% for floor area and 50\% for lot size and retain only unique best matches. This procedure yields 27,902 unique matches, corresponding to 44.9\% of housing permits in the Survey on Building Construction Started. After excluding observations with extreme values of lot size, the final estimation sample contains 23,144 observations.\footnote{In the Survey on Building Construction Started, lot size is top-coded at 1,000 square meters. We therefore exclude observations reported as 1,000 square meters or more, since their exact parcel size is not observed.} The resulting matches are of high quality along the observed dimensions used in the procedure. By construction, the four discrete attributes match exactly. For the continuous variables, the median percentage difference is 0.8\% for total floor area and 0.0\% for lot size. Even at the 90th percentile, the differences remain modest, at 6.5\% for floor area and 6.0\% for lot size. These statistics indicate that the retained matches are close not only in discrete characteristics but also in the scale of the building and parcel.

The matched estimation sample is not fully representative of either source population. Relative to unmatched observations, matched observations are more likely to be apartment buildings and non-wood structures. This pattern reflects the matching procedure and the requirement of a unique best match. The empirical estimates should therefore be interpreted as applying to the matched new-construction sample in Tokyo's 23 special wards.

The quasi-panel construction is the empirical analogue of the Markov moments. Since the same parcel is generally not observed repeatedly, we approximate the lagged local state using geographically nearby observations from the previous period. Building centroid coordinates are available for the matched estimation sample, so the baseline lag links each period $t$ observation to a nearby period $t-1$ observation using centroid coordinates. To characterize the spatial proximity of matched pairs, Table \ref{tab:match_distance} reports the distribution of matching distances. For the full sample, the median distance between matched observations is 127 meters, with the 90th percentile at 288 meters. The distribution is wider in the single-family subsample, with a median of 191 meters and a 90th percentile of 440 meters. These distances describe the spatial scale of the local-homogeneity approximation used in the baseline implementation. They do not by themselves validate the approximation, so the robustness analysis also considers alternative geographic lag constructions, including distance-threshold and nearest-neighbor averages.

\begin{table}[htbp]
\centering
\caption{Distance between matched pairs in the quasi-panel}
\label{tab:match_distance}
\begin{tabular}{lccccc}
\toprule
 & Mean & Median & P25 & P75 & P90 \\
\midrule
Full sample & 150 & 127 & 74 & 199 & 288 \\
Single-family & 228 & 191 & 108 & 304 & 440 \\
\bottomrule
\end{tabular}
\begin{minipage}{0.9\textwidth}\scriptsize
    \textit{Notes:} Distances are measured in meters using building centroid coordinates.
\end{minipage}
\end{table}

Before turning to descriptive statistics, we detrend capital and land prices to make them comparable across years. Following \citet{Combes2021-pu}, we regress $k_{jt} \equiv \log K_{jt}$ and $\log p^l_{jt}$ on year fixed effects and use the residuals in all subsequent analyses. Land acquisition cost $R_{jt}$ and housing value $V_{jt}$ are then recomputed from detrended capital and land prices. This detrending serves two purposes. First, it makes the production-function variables comparable across years. Second, it aligns the empirical setting with the Markov structure on $\tilde{\omega}_{jt}$ assumed in Assumption \ref{as:markov}, which abstracts from builders' expectations of future aggregate conditions.

Table \ref{tab:descriptive} reports descriptive statistics for the estimation sample. The full sample contains 23,144 observations from Tokyo's 23 special wards, of which 10,190 (44.0\%) are single-family housing. To characterize the input variation available for estimation, we report both central tendencies and 10th to 90th percentile ranges. In the full sample, the average lot size is 273.8 m$^2$, with 10th and 90th percentiles of 75.6 and 631.9 m$^2$. Average construction cost is 8,133 and average land acquisition cost is 8,214, both in units of 10,000 yen. The corresponding 10th to 90th percentile ranges are 2,547 to 16,544 and 1,830 to 18,125, respectively. The implied average housing value is 16,347 and the average capital share is 0.519. To assess how the input mix differs across housing types, we also report statistics for the single-family subsample. The average lot size is 284.4 m$^2$, average construction cost is 4,671, and average land acquisition cost is 8,343. The implied average housing value is 13,014 and the average capital share is 0.451.

\begin{table}[htbp]
\centering
\caption{Descriptive Statistics}
\label{tab:descriptive}
\small
\begin{tabular}{lrrrrr}
\toprule
 & Mean & Median & SD & P10 & P90 \\
\hline
\multicolumn{6}{l}{\textit{Panel A: Full Sample (N = 23,144)}} \\ [3pt]
Lot size $L$ (m$^2$) & 273.8 & 191.7 & 221.9 & 75.6 & 631.9 \\
Construction cost $K$ (\textyen 10K) & 8,133 & 5,501 & 8,050 & 2,547 & 16,544 \\
Land acquisition cost $R$ (\textyen 10K) & 8,214 & 5,391 & 8,394 & 1,830 & 18,125 \\
Unit land price $p_l$ (\textyen 1K/m$^2$) & 312 & 265 & 248 & 167 & 461 \\
Capital share $K/V^{zp}$ & 0.519 & 0.550 & --- & --- & --- \\
\\
\multicolumn{6}{l}{\textit{Panel B: Single-Family Housing (N = 10,190)}} \\ [3pt]
Lot size $L$ (m$^2$) & 284.4 & 175.6 & 250.3 & 64.2 & 709.7 \\
Construction cost $K$ (\textyen 10K) & 4,671 & 4,019 & 2,892 & 2,143 & 7,673 \\
Land acquisition cost $R$ (\textyen 10K) & 8,343 & 4,907 & 9,134 & 1,557 & 19,361 \\
Unit land price $p_l$ (\textyen 1K/m$^2$) & 308 & 256 & 274 & 162 & 461 \\
Capital share $K/V^{zp}$ & 0.451 & 0.474 & --- & --- & --- \\
\bottomrule
\end{tabular}
\begin{minipage}{0.9\textwidth}\scriptsize
    \textit{Notes:} Data are from the matched building starts, current land use, and land price surveys for Tokyo's 23 special wards, 2012--2016. Construction cost, approximated land acquisition cost, and the zero-profit-implied value measure are reported in units of 10,000 yen. Unit land price is reported in units of 1,000 yen per square meter. The capital share is computed as $K_{jt}/V^{zp}_{jt}$. Construction cost, unit land price, and derived variables are detrended for year effects as described above.
\end{minipage}
\vspace{0.1cm}
\end{table}

\subsection{Results}
Table~\ref{tab:estimates} reports the estimated output elasticities for capital and land in the matched estimation sample and in the single-family subsample. The estimates compare the proposed method, which explicitly models latent local conditions, with the CDG benchmark. The results should be interpreted under the maintained zero-profit-implied value construction described above. Three findings stand out. First, the proposed estimator yields larger capital elasticities than the CDG benchmark in both samples. Second, the single-family subsample exhibits a lower capital elasticity and a higher land elasticity than the full matched sample. Third, because the proposed estimator recovers both capital and land elasticities, returns to scale can be assessed rather than imposed; the estimated returns to scale are close to one under the maintained value construction.

The first result is that explicitly modeling latent local conditions matters quantitatively for the recovered capital elasticity in this application. In the full matched sample, the proposed estimator yields a capital elasticity of $0.510$, whereas the CDG benchmark yields $0.453$. In the single-family subsample, the corresponding estimates are $0.440$ and $0.337$. These differences are consistent with the transmission-bias mechanism highlighted in the Monte Carlo results. When capital input responds to the unobserved composite state, failing to account for that state can distort the recovered capital elasticity. The direction of this bias depends on the underlying data-generating process and is not determined a priori, so the empirical estimates need not mirror the sign of the bias observed in a particular Monte Carlo design.

The second result is that the single-family subsample displays a different input mix from the full matched sample. Under the proposed estimator, the land elasticity is $0.486$ in the full matched sample and $0.563$ in the single-family subsample, while the capital elasticity falls from $0.510$ to $0.440$. This pattern suggests that, within the matched estimation sample, single-family housing relies more heavily on land and less on structure than the broader sample, which includes more capital-intensive multi-family construction. The gap between the proposed and CDG capital elasticities is also larger in the single-family subsample: $0.058$ in the full matched sample and $0.103$ in the single-family subsample.

The third result concerns returns to scale. Because the proposed estimator also recovers the land elasticity, returns to scale can be assessed rather than imposed. In the full matched sample, the sum of the estimated capital and land elasticities is $0.996$. In the single-family subsample, the sum is $1.004$. Thus, constant returns are not imposed by the estimator, although the point estimates are close to one under the maintained value construction.

The empirical results should be interpreted with two implementation issues in mind. First, the value measure used in estimation is the zero-profit-implied object $V^{zp}_{jt}$ rather than an observed transaction value. Second, the Markov moments are implemented using a quasi-panel lag constructed from nearby previous-period observations rather than repeated observations of the same parcel. Appendix~\ref{app:empirical_robustness} reports robustness exercises for both issues. The value-construction exercises examine how the recovered elasticities change under alternative constructed value measures, and the lag-construction exercises compare the baseline nearest-neighbor quasi-panel estimates with administrative-cell, averaged-lag, and placebo alternatives.

\begin{table}[htbp]
\centering
\caption{Production Function Estimates: Comparison of Methods}
\label{tab:estimates}
\small
\begin{tabular}{lcccccc}
\toprule
 & \multicolumn{3}{c}{Full Sample} & \multicolumn{3}{c}{Single-Family} \\
 \cmidrule(lr){2-4} \cmidrule(lr){5-7}
 & Estimate & SE & 95\% CI & Estimate & SE & 95\% CI \\
\hline
\multicolumn{7}{l}{\textit{Panel A: CDG (Combes, Duranton, Gobillon 2021)}} \\[3pt]
$\widehat{\bar{D}}_k$ & 0.452 & 0.014 & $[0.425,\ 0.480]$ & 0.337 & 0.013 & $[0.312,\ 0.362]$ \\
\\
\multicolumn{7}{l}{\textit{Panel B: Proposed method}} \\[3pt]
$\widehat{\bar{D}}_k$ & 0.510 & 0.012 & $[0.487,\ 0.534]$ & 0.440 & 0.013 & $[0.415,\ 0.466]$ \\
$\widehat{\bar{D}}_l$ & 0.486 & 0.011 & $[0.463,\ 0.508]$ & 0.563 & 0.013 & $[0.537,\ 0.589]$ \\
$\widehat{\bar{D}}_k + \widehat{\bar{D}}_l$ & 0.996 & 0.004 & $[0.989,\ 1.003]$ & 1.004 & 0.006 & $[0.993,\ 1.015]$ \\
\bottomrule
\end{tabular}
\begin{minipage}{0.9\textwidth}\scriptsize
    \textit{Notes:}  Full sample: $N = 23{,}144$; single-family subsample: $N = 10{,}190$ (44.0\%).  $k$ and $l$ denote capital and land. SEs are ward-cluster bootstrap SEs with $B=300$.  95\% CI = estimate $\pm$ 1.96 $\times$ SE.  All estimates use year-demeaned variables following \citet{Combes2021-pu}. Panel~B uses nearest-neighbor quasi-panel lags constructed from building centroid coordinates. The returns-to-scale SE is computed as the bootstrap standard deviation of $\widehat{\bar{D}}_k+\widehat{\bar{D}}_l$.
\end{minipage}
\end{table}

\section{Conclusion}\label{sec:conclusion}
This paper develops an estimation method for the housing production function that treats unobserved location-level heterogeneity as a central identification problem rather than as a residual disturbance. A central difficulty in estimating housing production is that builders choose inputs in response to local conditions that are observed by them but not by the econometrician. When this payoff-relevant heterogeneity affects both revenue and input choice, existing approaches can fail to separate technological relationships from endogenous responses to latent profitability. 
To address this problem, this paper combines the capital share equation with a scalar Markov restriction on the payoff-relevant local state. The share equation identifies the capital elasticity because the latent state drops out of the first-order condition, while the Markov moments recover the land-specific integration component and the state transition process. In the empirical application, housing value is not directly observed, so the estimator is applied to a zero-profit-implied value object constructed from capital and land costs.

The Monte Carlo simulations show that the proposed estimator recovers the relevant elasticities when the maintained scalar-state and Markov restrictions hold. This result is important because the identification problem is most severe when production elasticities vary across observations and capital choices respond to latent local conditions. In the Cobb--Douglas designs, both the proposed estimator and the CDG benchmark recover the capital elasticity well. In the CES designs, where the elasticity varies with the input mix, the proposed estimator recovers both capital and land elasticities, while the benchmark can transmit part of the latent-state response into the recovered production function. The decreasing-returns designs also show that the estimator does not mechanically impose constant returns to scale. These exercises support the main identification logic while making clear that the estimator's performance is evaluated under the maintained state and timing restrictions.

The empirical application illustrates how the estimator can be implemented in a dense single-city setting. We apply the method to newly constructed housing in Tokyo's 23 special wards, combining administrative construction records, land-use survey data, and roadside land price data. Because housing value is not directly observed, the estimation uses a zero-profit-implied value object constructed from capital and land costs. Within the matched estimation sample, the proposed estimator yields larger capital elasticities than the CDG benchmark, with a larger gap in the single-family subsample. The method also recovers land elasticities, so returns to scale can be assessed rather than imposed. Under the maintained value construction, the estimated returns to scale are close to one in both the full matched sample and the single-family subsample. The empirical results therefore show that explicitly modeling latent local conditions can matter quantitatively for the recovered capital-land elasticities.

Several directions for future research remain. First, the empirical application in this paper is necessarily constrained by data availability. In particular, land acquisition cost and housing value are not directly observed in a single integrated dataset and must instead be approximated or recovered from auxiliary information and the zero-profit relationship. Although this strategy is consistent with the structure of the model and performs well in the Monte Carlo exercises, richer data containing realized transaction values, more precise land acquisition costs, and parcel-level panel information would permit a sharper assessment of measurement error and identification. Second, the empirical analysis is conducted for newly constructed housing in Tokyo's 23 special wards, which constitute a dense and highly regulated urban environment. Extending the analysis to suburban areas, smaller cities, or other metropolitan regions would clarify how far the estimated elasticities generalize across different institutional and spatial settings, and would also permit a direct methodological comparison with approaches that rely on cross-urban variation, such as \citet{Combes2021-pu}. Third, the latent state in the present framework is interpreted as a reduced-form location-level profitability component relevant for builders' decisions. A useful next step would be to connect this latent heterogeneity more directly to observable determinants such as land-use regulation, topography, infrastructure access, and neighborhood demand conditions. Finally, because the estimated production function is recovered under the prevailing regulatory regime, an important extension would be to use the framework for counterfactual analysis of zoning, floor-area-ratio restrictions, and other land-use policies. Such analysis should allow for the possibility that builders' effective production technology and input responses may themselves change when the policy environment changes.

\clearpage

\begin{spacing}{1.0}
\ifx\undefined\bysame
\newcommand{\bysame}{\leavevmode\hbox to\leftmargin{\hrulefill\,\,}}
\fi

\end{spacing}

\clearpage

\section*{Appendix}

\setcounter{table}{0}
\renewcommand{\thetable}{A\arabic{table}}
\setcounter{equation}{0}
\renewcommand{\theequation}{A\arabic{equation}}
\setcounter{section}{0}
\renewcommand{\thesection}{\Alph{section}}

\section{Model appendix}
\subsection{Finite-Dimensional Implementation} \label{app:practical_estimator}
This appendix describes the implementable version of the estimator used in the simulations and empirical application. The goal is to translate the identification argument in the main text into a finite-dimensional procedure. We adapt the framework of \citet{Gandhi2020-kx} and \citet{Chen2007-hi} to a cross-sectional setting with location-level heterogeneity. The procedure consists of (i) estimation of the capital cost share equation using a truncated polynomial series, followed by analytical integration to recover the production function up to an integration component, and (ii) recovery of the land-specific integration component $C(l)$ and the Markov transition function $g$ via GMM.

\subsubsection*{Step 1: Estimation of the Share Equation and Integration}
We approximate the capital cost share function using a finite-dimensional polynomial series in log capital and log land inputs. Let $k_{jt} = \ln K_{jt}$, $l_{jt} = \ln L_{jt}$, and $v_{jt} = \ln V_{jt}$ denote log capital, log land, and log housing value for the observation in location $j$ at time $t$, with $j = 1, \ldots, J$ and $t = 1, \ldots, T$. Let $V_{jt}$ denote the value measure supplied to the estimator. When housing value is directly observed, $V_{jt}=P^h_{jt}Y_{jt}$. In the empirical application, housing value is not directly observed, so the value measure is the zero-profit-implied object $ V^{zp}_{jt}=K_{jt}+R_{jt}$. For notational simplicity, we write $V_{jt}$ in the estimator, with the understanding that $V_{jt}=V^{zp}_{jt}$ in the Tokyo application.

The capital cost share $s^k_{jt} \equiv K_{jt}/V_{jt}$ satisfies the share equation derived from the production function:
\begin{equation*}
\ln s^k_{jt} = \ln D(k_{jt}, l_{jt}) - \epsilon_{jt},
\end{equation*}
where $D(k, l) \equiv \partial h/\partial k$ is the elasticity of output with respect to capital, and $\epsilon_{jt}$ is the idiosyncratic shock defined in the main text. We approximate the capital elasticity function $D(k,l)$ by a complete polynomial of degree $r$:
\begin{equation*}
D(k,l) = \sum_{r_k+r_l\leq r} \gamma_{r_k r_l} k^{r_k}l^{r_l}.
\end{equation*}
The coefficients are estimated from the share equation by nonlinear least squares:
\begin{equation*}
\widehat{\gamma} = \arg\min_{\gamma} \frac{1}{n} \sum_{j,t} \left[ \log s^k_{jt} - \log D(k_{jt},l_{jt};\gamma)
\right]^2.
\end{equation*}
In the empirical application, we use a second-degree polynomial ($r = 2$), so the capital elasticity specification takes the form
\begin{equation*}
D(k_{jt}, l_{jt}) = \gamma_0 + \gamma_k k_{jt} + \gamma_l l_{jt} + \gamma_{kk} k_{jt}^2 + \gamma_{kl} k_{jt} l_{jt} + \gamma_{ll} l_{jt}^2.
\end{equation*}
Because the share equation takes the logarithm of the elasticity approximation, the implementation restricts the fitted elasticity $D(k,l)$ to be positive over the estimation sample.

Given the estimated parameters, we analytically integrate $\hat{D}$ with respect to $k$ to recover the production function up to an integration component:
\begin{equation*}
\int \hat{D}(k_{jt}, l_{jt}) \, dk_{jt} = \sum_{\substack{r_k + r_l \le r \\ r_k, r_l \ge 0}} \frac{\hat{\gamma}_{r_k r_l}}{r_k + 1} k_{jt}^{r_k + 1} l_{jt}^{r_l}.
\end{equation*}
Using the estimated share equation residual $\hat{\varepsilon}_{jt}$, we construct
\begin{equation*}
\hat{\phi}_{jt} \equiv v_{jt} - \hat{\varepsilon}_{jt} - \int \hat{D}(k_{jt}, l_{jt}) \, dk_{jt}.
\end{equation*}
As discussed in the estimation strategy, the quantity $\hat{\phi}_{jt}$ absorbs the integration component $C(l_{jt})$ and the composite state $\tilde{\omega}_{jt} \equiv \omega_{jt} + p^h_{jt}$, which combines the location-level productivity $\omega_{jt}$ with the location-time component of log output prices $p^h_{jt}$.

\subsubsection*{Step 2: Recovery of the Integration Component and Transition Process}
We next recover the land-specific integration component $C(l)$ and the Markov transition function $g$ using GMM. The integration component $C(l)$ is a deterministic function of land input and reflects features of the production technology that cannot be separately identified from variation in $k$ alone. We specify it as a second-degree polynomial in log land:
\begin{equation*}
C(l_{jt};\alpha) = \alpha_0 + \alpha_l l_{jt} + \alpha_{ll} l_{jt}^2.
\end{equation*}
The intercept in $C(l;\alpha)$ is included in the numerical implementation even though the level of $C(l)$ cannot be separately identified from the mean of the recovered composite state. For any scalar $a$, replacing $C(l)$ with $C(l)+a$ and $\tilde{\omega}_{jt}$ with $\tilde{\omega}_{jt}-a$ implies the same adjusted value. This normalization does not affect the capital elasticity, the land elasticity, or returns to scale, because these objects depend on derivatives of $h(k,l)$. We therefore estimate the intercept jointly with the slope and curvature terms and recenter the recovered state after estimation.

The composite state is assumed to follow a first-order Markov process:
\begin{equation*}
\tilde{\omega}_{jt} = g(\tilde{\omega}_{j,t-1}) + \eta_{jt},
\end{equation*}
where $\eta_{jt}$ is a mean-zero innovation satisfying $\mathbb{E}[\eta_{jt} \mid I_{j,t-1}] = 0$. This Markov structure, together with the timing assumption that the lagged instruments are predetermined with respect to $\eta_{jt}$, provides the basis for the Step~2 moment restrictions. Since $g$ is not known in closed form, we approximate it by a cubic polynomial:
\begin{equation*}
g(\tilde{\omega}_{j,t-1}) = \sum_{a=0}^{3} \delta_a \tilde{\omega}_{j,t-1}^a.
\end{equation*}
A cubic polynomial offers sufficient flexibility to capture potential nonlinearities in the transition process while keeping the dimensionality of the parameter space manageable. Monte Carlo simulations reported in Section~\ref{sec:monte_carlo} confirm that this specification performs well across a range of data-generating processes.

Because the same parcel is generally not observed repeatedly, the data form a quasi-panel rather than a true parcel panel. The empirical implementation of the composite state is based on the Markov structure that links $\tilde{\omega}_{jt}$ across nearby locations over time, not on repeated observations of the same physical parcel.\footnote{If multiple constructions were observed in the same location-time cell, one could average the proxies $\hat{\phi}_{jt} - C(l_{jt};\alpha)$ within the cell. The implementation used here instead treats each observed construction as a location-time observation and uses matched lagged observations.}
For a candidate parameter vector $\alpha = (\alpha_0, \alpha_l, \alpha_{ll})$:
\begin{enumerate}
    \item Construct the proxy for the composite state:
    \begin{equation*}
    \hat{\omega}_{jt}(\alpha) = \hat{\phi}_{jt} - C(l_{jt};\alpha).
    \end{equation*}
    \item Estimate the auxiliary transition model using the matched previous-period observation:
    \begin{equation*}
    \hat{\omega}_{jt}(\alpha) = \sum_{a=0}^{3} \delta_a \hat{\omega}_{m(j,t)}(\alpha)^a + u_{jt},
    \end{equation*}
    and obtain the residuals $\hat{u}_{jt}$.
    \item Evaluate the GMM objective using moment conditions based on lagged land inputs:
    \begin{equation*}
    Q(\alpha) = \left\| \frac{1}{n} \sum_{j,t} \hat{u}_{jt} \cdot z_{m(j,t)} \right\|^2, \quad z_{m(j,t)} = \begin{pmatrix} 1 \\ l_{m(j,t)} \\ l_{m(j,t)}^2 \end{pmatrix}.
    \end{equation*}
\end{enumerate}

Given the estimated functions, the land elasticity is computed as
\begin{equation*}
D_l(k,l) = \frac{\partial}{\partial l} \left[ \int^k \widehat{D}(\kappa,l)\,d\kappa \right] + C'(l;\widehat{\alpha}).
\end{equation*}

\subsection{Spatial correlation in the latent state}
\label{app:spatial_correlation}

This subsection discusses how spatial correlation in the composite state affects the proposed estimator. The Markov restriction in the main text governs the time-series evolution of the payoff-relevant state. Spatial dependence may still remain in the cross-sectional distribution of the state, in the innovations, or in the resulting moment conditions across locations. The main point is that spatial dependence changes the joint distribution of the sample moments and the asymptotic variance, but it does not by itself invalidate the conditional moment restrictions used by the estimator, provided the moment process satisfies a spatial law of large numbers. This distinction follows the standard treatment of GMM and extremum estimators under dependent observations \citep{Hansen1982,Newey1994-mt} and the spatial asymptotic results for random fields in \citet{Jenish2009-ab}.

A simple way to represent spatial dependence is to allow the composite state to contain a spatially correlated component. For example, at a given date, the vector of composite states may satisfy a spatial autoregressive representation,
\begin{equation*}
\tilde{\omega}_t = \rho W\tilde{\omega}_t + \eta_t,
\end{equation*}
where $W$ is a row-standardized spatial weights matrix and $|\rho|<1/\lambda_{\max}(W)$. This representation is only an example of cross-sectional spatial dependence, as in standard spatial econometric models \citep{Anselin1988,Kelejian1999}. The consistency argument below does not rely on this particular spatial autoregressive specification. Instead, it uses high-level weak-dependence conditions for the moment process generated by the estimator.

The distinction between moment validity and dependence is important. For the share equation in Step 1, the relevant restriction is the timing assumption that the ex post revenue shock is realized after the capital choice. Under this timing assumption,
\begin{equation*}
\mathbb{E}[\epsilon_{jt}\mid k_{jt},l_{jt}]=0.
\end{equation*}
Spatial correlation in the latent state affects the joint distribution of observations across locations, but it does not change this conditional mean restriction. For the Markov moments in Step 2, the relevant restriction is
\begin{equation*}
\mathbb{E}[\eta_{jt}\mid I_{j,t-1}]=0,
\end{equation*}
together with the use of variables measured before the current innovation as instruments. Spatial dependence may induce correlation between moment conditions across nearby locations, but it does not change the population orthogonality condition as long as the instruments are predetermined with respect to the current innovation.

The empirical quasi-panel construction introduces an additional approximation. In the data, the same parcel is generally not observed repeatedly, so the lagged state is proxied by nearby previous-period observations. The argument in this subsection concerns spatial dependence in the moment process. It does not by itself establish that the nearest-neighbor lag perfectly recovers the true lagged state. That approximation is evaluated separately through the quasi-panel robustness checks in the empirical appendix and through the Monte Carlo exercises with alternative lag constructions.

We state the high-level regularity conditions as follows. Let $\theta$ denote the finite-dimensional parameter vector used in the practical estimator after imposing the level normalization discussed in Appendix~\ref{app:practical_estimator}. Let $\theta_0$ denote the corresponding target parameter.

\begin{assumption}[High-level regularity conditions under spatial dependence]
\label{as:spatial_regular}
The following conditions hold.
\begin{enumerate}
    \item The parameter space $\Theta$ is compact, and $\theta_0$ belongs to the interior of $\Theta$.
    \item The population objective function $Q(\theta)$ is continuous on $\Theta$ and is uniquely minimized at $\theta_0$.
    \item The sample objective function $Q_n(\theta)$ converges uniformly in probability to $Q(\theta)$ over $\Theta$.
    \item The spatial dependence in the moment process is weak enough for a spatial law of large numbers to hold.
    \item The moment functions are uniformly bounded in a neighborhood of $\theta_0$, or are dominated by an integrable envelope function.
    \item The weighting matrix $\widehat{W}$ converges in probability to a positive definite matrix $W$.
\end{enumerate}
\end{assumption}

These conditions are stated at a high level. The purpose is not to provide primitive conditions for every possible form of spatial dependence. Rather, the conditions clarify that the probability limit of the estimator is unchanged when spatial dependence is weak enough for the sample moments to converge uniformly to their population counterparts. This is the standard structure of a high-level consistency argument for GMM and extremum estimators \citep{Hansen1982,Newey1994-mt}, with the additional requirement that the spatially dependent moment array satisfies an appropriate law of large numbers \citep{Jenish2009-ab}.

Let $m_{jt}(\theta)$ denote the stacked moment vector used in the practical estimator. The sample average moment is
\begin{equation*}
\bar{m}_n(\theta) = \frac{1}{n} \sum_{j,t} m_{jt}(\theta),
\end{equation*}
and the GMM objective is
\begin{equation*}
Q_n(\theta) = \bar{m}_n(\theta)^{\prime} \widehat{W} \bar{m}_n(\theta).
\end{equation*}

\begin{proposition}[Consistency under high-level spatial-dependence conditions]
\label{prop:spatial_consistency}
Under Assumption~\ref{as:spatial_regular}, the GMM estimator
\begin{equation*}
\widehat{\theta} = \arg\min_{\theta\in\Theta} Q_n(\theta)
\end{equation*}
satisfies
\begin{equation*}
\widehat{\theta} \overset{p}{\to} \theta_0
\end{equation*}
as the cross-sectional dimension grows.
\end{proposition}

\begin{proof}
The timing assumptions in the main text imply the population moment restrictions at $\theta_0$. For the Step 1 share equation, the ex post revenue shock is realized after the capital choice, so the share-equation residual has mean zero conditional on the inputs. For the Step 2 Markov moments, the innovation $\eta_{jt}$ is mean independent of the information available at the end of the previous period, and the instruments are functions of variables measured before the current innovation. Therefore,
\begin{equation*}
\mathbb{E}[m_{jt}(\theta_0)]=0.
\end{equation*}
Spatial correlation affects the covariance structure of $m_{jt}(\theta_0)$ across locations, but not its population mean.

By Assumption~\ref{as:spatial_regular}, the sample objective $Q_n(\theta)$ converges uniformly in probability to the population objective $Q(\theta)$. The population objective is continuous and uniquely minimized at $\theta_0$. Standard extremum-estimator arguments then imply
\begin{equation*}
\widehat{\theta}
\overset{p}{\to}
\theta_0.
\end{equation*}
\end{proof}

Spatial dependence is more consequential for inference than for the population moment restrictions. Under cross-sectional independence, the asymptotic variance of the sample moments depends only on the variance of the individual moment contributions. Under spatial dependence, the covariance between nearby locations also contributes. If a central limit theorem for weakly spatially dependent moments holds, then
\begin{equation*}
\sqrt{n}\,\bar{m}_n(\theta_0) \overset{d}{\to} N(0,\Omega_{sp}),
\end{equation*}
where
\begin{equation*}
\Omega_{sp} = \lim_{n\to\infty} \operatorname{Var} \left( \frac{1}{\sqrt{n}} \sum_{j,t} m_{jt}(\theta_0) \right).
\end{equation*}
The matrix $\Omega_{sp}$ includes spatial covariance terms. Ignoring these terms can understate sampling uncertainty when nearby observations have correlated latent states or correlated Markov innovations. This is the reason spatial-HAC and related estimators are commonly used for cross-sectional dependence in moment-based estimation \citep{Conley1999-ck,Kelejian2007}.

The corresponding asymptotic variance of the GMM estimator takes the usual sandwich form,
\begin{equation*}
\left(G^{\prime}WG\right)^{-1} G^{\prime}W \Omega_{sp} WG \left(G^{\prime}WG\right)^{-1}, 
\end{equation*}
where
\begin{equation*}
G = \mathbb{E} \left[ \frac{\partial m_{jt}(\theta_0)}{\partial \theta^{\prime}} \right].
\end{equation*}
Thus, spatial dependence does not require a different population moment condition, but it does require inference that accounts for cross-location dependence. In the empirical application, we therefore interpret conventional standard errors cautiously and use resampling and robustness exercises to examine the sensitivity of the estimates to spatial and quasi-panel implementation choices.

The Monte Carlo exercises in Appendix~\ref{app:mc_spatial} complement this high-level argument. They allow the latent state to contain spatial correlation and show that the baseline Monte Carlo conclusions are not driven by an independent cross-sectional latent-state design. Those simulations should be interpreted as finite-sample evidence, not as a proof that the maintained assumptions hold in the empirical application.

\setcounter{table}{0}
\renewcommand{\thetable}{B\arabic{table}}
\setcounter{equation}{0}
\renewcommand{\theequation}{B\arabic{equation}}

\section{Bias in Existing Estimators}
\label{app:bias_existing}

This section explains why ignoring unobserved location-level heterogeneity can generate transmission bias in two existing frameworks for estimating the housing production function, \citet{Combes2021-pu} (CDG) and \citet{Epple2010-rg} (EGS). As in the main text, we suppress a separate construction-level index and treat each observed construction as a location-time observation $(j,t)$. Let total housing value be denoted by $V_{jt} \equiv P^h_{jt}Y_{jt}$. The housing price $P^h_{jt}$ and physical housing services $Y_{jt}$ are not separately observed. For simplicity, this appendix writes the latent local component as $\omega_{jt}$. It can be interpreted as the payoff-relevant local component that enters the revenue product of capital, corresponding to the composite state in the main text.

\subsection{Bias in Combes, Duranton, and Gobillon (2021)}

The CDG approach recovers the housing production function by integrating the capital output elasticity implied by the first-order condition for capital. The potential difficulty is that capital is chosen after builders observe local conditions. When the latent local component affects the expected return to capital, observed capital variation reflects both technological substitution and builders' endogenous responses to this component.

Suppose housing services are produced according to
\begin{equation*}
Y_{jt} = H(K_{jt},L_{jt})\exp(\omega_{jt}).
\end{equation*}
A builder chooses capital under perfect competition, taking the housing price $P^h_{jt}$, land input $L_{jt}$, and the capital price $P^k=1$ as given. The first-order condition for capital is
\begin{equation*}
P^h_{jt} \frac{\partial H(K_{jt},L_{jt})}{\partial K_{jt}} \exp(\omega_{jt}) = 1. 
\end{equation*}
This condition determines the optimal capital choice as a function of the housing price, land input, and the latent local component:
\begin{equation*}
K_{jt} = K^*(P^h_{jt},L_{jt},\omega_{jt}).
\end{equation*}

Define the output elasticity of capital as
\begin{equation*}
D(K_{jt},L_{jt}) \equiv \frac{\partial \log H(K_{jt},L_{jt})}{\partial \log K_{jt}}.
\end{equation*}
At the builder's optimum, capital is chosen after observing the latent local component, so the observed capital input can be written as
\begin{equation*}
K_{jt} = K^*(P^h_{jt},L_{jt},\omega_{jt}).
\end{equation*}
Multiplying the first-order condition by $K_{jt}$ and dividing by housing value yields the capital share equation evaluated at this chosen input:
\begin{equation*}
D(K^*(P^h_{jt},L_{jt},\omega_{jt}),L_{jt}) = \frac{K^*(P^h_{jt},L_{jt},\omega_{jt})}{V_{jt}}.
\end{equation*}
Using the zero-profit accounting relationship,
\begin{equation*}
V_{jt} = K^*(P^h_{jt},L_{jt},\omega_{jt}) + R(P^h_{jt},L_{jt},\omega_{jt}),
\end{equation*}
this can be written as
\begin{equation}
D(K^*(P^h_{jt},L_{jt},\omega_{jt}),L_{jt}) = \frac{K^*(P^h_{jt},L_{jt},\omega_{jt})}{K^*(P^h_{jt},L_{jt},\omega_{jt}) + R(P^h_{jt},L_{jt},\omega_{jt})}.
\label{eq:elasticity_integrand_appendix}
\end{equation}

Equation~\eqref{eq:elasticity_integrand_appendix} shows the source of the problem. The latent component $\omega_{jt}$ drops out of the share equation when the equation is evaluated at the realized input choices, but it remains in the equilibrium choices that generate the observed capital share. The observed integrand is therefore evaluated along a locus of choices indexed by $\omega_{jt}$. If $\omega_{jt}$ is ignored, the right-hand side is treated as if it were a single-valued function of $(K,L)$ alone, even though variation in $K$ conditional on $L$ may partly reflect variation in the latent local component. In addition, land costs may vary with $\omega_{jt}$ even after conditioning on observed inputs. Recovering the production function requires varying capital while holding fixed the latent local component, but this component is unobserved. Integrating equation~\eqref{eq:elasticity_integrand_appendix} while ignoring $\omega_{jt}$ therefore mixes the partial effect of capital on $\log H$ with the endogenous response of capital and land costs to the latent local component. In general, the resulting integral does not recover the true production function $H$.\footnote{This issue is conceptually related to the transmission bias discussed by \citet{Gandhi2020-kx} in the context of production function estimation with flexible inputs. The specific mechanism here differs because the bias arises from the non-uniqueness of the observed capital-share integrand as a function of $(K,L)$ when latent local conditions affect both capital choice and land costs.}

\subsection{Bias in Epple, Gordon, and Sieg (2010)}

The EGS approach uses duality under constant returns to scale (CRS) to recover the production function from land prices and housing values per unit of land. Let $P^L_{jt}$ denote the land price per unit of land in location $j$ at time $t$, and let $V^L_{jt} \equiv \frac{V_{jt}}{L_{jt}}$ denote housing value per unit of land.

Under CRS, the production function can be written per unit of land as
\begin{equation*}
H(K_{jt},L_{jt}) = L_{jt}q(K^L_{jt},1), 
\end{equation*}
where $K^L_{jt} \equiv \frac{K_{jt}}{L_{jt}}$ is the capital--land ratio. Under perfect competition, the first-order condition for capital is
\begin{equation*}
P^h_{jt} \frac{\partial q(K^L_{jt},1)}{\partial K^L_{jt}} \exp(\omega_{jt}) = 1.
\end{equation*}
This condition implicitly defines the optimal capital--land ratio as
\begin{equation*}
K^L_{jt} = \kappa(P^h_{jt},\exp(\omega_{jt})).
\end{equation*}
The corresponding housing services per unit of land are
\begin{equation*}
S(P^h_{jt},\exp(\omega_{jt})) \equiv q(\kappa(P^h_{jt},\exp(\omega_{jt})),1).
\end{equation*}

Because the housing price $P^h_{jt}$ is unobserved, EGS exploit the fact that, for a given value of $\omega_{jt}$, housing value per unit of land satisfies
\begin{equation*}
V^L_{jt} = P^h_{jt} S(P^h_{jt},\exp(\omega_{jt})) \exp(\omega_{jt}).
\end{equation*}
Under the relevant monotonicity condition, this relationship can be inverted to obtain
\begin{equation*}
P^h_{jt} = P^h(V^L_{jt},\exp(\omega_{jt})).
\end{equation*}
The zero-profit condition per unit of land then implies
\begin{equation*}
P^L_{jt} = V^L_{jt} - \kappa(P^h(V^L_{jt},\exp(\omega_{jt})),\exp(\omega_{jt})).
\end{equation*}
When $\omega_{jt}$ is homogeneous across locations, this equation establishes a monotonic relationship between the observed land price and value per unit of land:
\begin{equation*}
P^L_{jt} = r(V^L_{jt}).
\end{equation*}
\citet{Epple2010-rg} show that this equilibrium mapping, together with the envelope theorem, identifies the supply function under CRS.

The difficulty arises when $\omega_{jt}$ varies across locations. The mapping from $V^L_{jt}$ to $P^L_{jt}$ is no longer a function of $V^L_{jt}$ alone. It also depends on $\omega_{jt}$ through the inversion for $P^h_{jt}$ and through the capital--land ratio $\kappa(\cdot)$. Two locations with the same $V^L_{jt}$ can have different $P^L_{jt}$ because they correspond to different values of $\omega_{jt}$. Estimating $r(\cdot)$ from observed $(V^L_{jt},P^L_{jt})$ pairs while ignoring $\omega_{jt}$ therefore yields a function that does not equal the true equilibrium mapping. The recovered supply function need not equal the true supply function in general.\footnote{As in the CDG case, this issue is conceptually related to the transmission bias discussed by \citet{Gandhi2020-kx}.}

The EGS approach also relies critically on the CRS assumption. When CRS fails, for example under decreasing returns to scale, output per unit of land no longer reduces to a function of the capital--land ratio alone. The envelope-theorem argument underlying Proposition~1 of \citet{Epple2010-rg} therefore no longer applies in the same form, and the duality-based inversion does not identify the production function. In that case, the approach does not recover the production function even under homogeneous latent local conditions.

\setcounter{table}{0}
\renewcommand{\thetable}{C\arabic{table}}
\setcounter{equation}{0}
\renewcommand{\theequation}{C\arabic{equation}}

\section{Monte Carlo Simulation Results}\label{app:mc_additional}
This section reports additional Monte Carlo simulation results that complement the baseline analysis in the main text.

\subsection{Sensitivity to sample size}\label{app:mc_sample_size}

Table~\ref{tab:gnr_sensitivity} examines the finite-sample behavior of the proposed estimator as the number of regions $J$ varies from $10$ to $1{,}000$, holding one observed construction per region-period and $T=5$ periods fixed. The data-generating process is the CES CRS design, so the target satisfies $\bar{D}_k+\bar{D}_l=1$.

The capital elasticity is stable across the sample-size grid. The mean of $\widehat{\bar{D}}_k$ remains close to the target even when $J$ is small, and the scaled standard deviation changes little with $J$. The finite-sample instability instead comes from the land elasticity. At $J=10$, the mean of $\widehat{\bar{D}}_l$ is distorted by a small number of explosive replications. Once $J$ reaches $50$, the mean is close to the target, and the scaled standard deviation stabilizes as the cross section becomes larger. The CRS diagnostic leads to the same interpretation. The mean of $\widehat{\bar{D}}_k+\widehat{\bar{D}}_l$ is distorted at $J=10$, but is close to one for larger samples. Thus, the small-sample problem in this design is a tail instability in the Step 2 land-elasticity estimate, rather than a systematic shift in the center of the estimator.

\begin{table}[htbp]
\centering
\caption{Small-Sample Diagnostics for the Proposed Estimator}
\label{tab:gnr_sensitivity}
\small
\begin{tabular}{r ccc ccc cc}
\toprule
 & \multicolumn{3}{c}{$\widehat{\bar{D}}_k$}
 & \multicolumn{3}{c}{$\widehat{\bar{D}}_l$}
 & \multicolumn{2}{c}{$\widehat{\bar{D}}_k+\widehat{\bar{D}}_l$} \\
\cmidrule(lr){2-4} \cmidrule(lr){5-7} \cmidrule(lr){8-9}
$J$
& Target & Mean & SD$\sqrt{n}$
& Target & Mean & SD$\sqrt{n}$
& Target & Mean \\
\midrule
10      & 0.442 & 0.442 & 0.164 & 0.558 & 34.677 & 4{,}828.682 & 1.000 & 35.118 \\
50      & 0.442 & 0.443 & 0.172 & 0.558 & 0.556  & 4.440     & 1.000 & 0.999 \\
100     & 0.442 & 0.442 & 0.171 & 0.558 & 0.569  & 3.521     & 1.000 & 1.010 \\
200     & 0.442 & 0.442 & 0.168 & 0.558 & 0.562  & 2.078     & 1.000 & 1.003 \\
300     & 0.442 & 0.442 & 0.178 & 0.558 & 0.558  & 2.010     & 1.000 & 1.000 \\
500     & 0.442 & 0.442 & 0.152 & 0.558 & 0.559  & 1.885     & 1.000 & 1.000 \\
1{,}000 & 0.442 & 0.442 & 0.159 & 0.558 & 0.557  & 1.889     & 1.000 & 0.999 \\
\bottomrule
\end{tabular}
\begin{minipage}{0.95\textwidth}\scriptsize
\textit{Notes:} The data-generating process is CES with CRS. Each location has one observed construction per period ($N=1$), and the panel length is $T=5$. Results are based on 500 Monte Carlo replications. The effective GMM sample size is $n_{\text{gmm}} = N \times J \times (T-1)=4J$. ``Target'' is the cross-replication mean of the realized true average elasticity. SD$\sqrt{n}$ is the simulation standard deviation of the replication-level mean elasticity scaled by $\sqrt{n_{\text{gmm}}}$. The last two columns report diagnostics for the CRS restriction $\widehat{\bar{D}}_k+\widehat{\bar{D}}_l=1$.
\end{minipage}
\end{table}

\subsection{Benchmark with unobserved capital or housing value}\label{app:mc_unobserved_variables}

This subsection examines a benchmark implementation in which one variable required by the estimator is not observed. The exercise is motivated by the empirical application, where housing value is not directly observed and the estimator is applied to a zero-profit-implied value object. To isolate the missing-variable issue from measurement error in the constructed object, the simulation treats either capital input or housing value as unobserved and constructs the missing variable from the same zero-profit accounting relationship used in the data-generating process.

This exercise is not a test of the zero-profit condition. It asks whether the two-step estimator remains accurate when a required variable is constructed from the maintained accounting relationship. Table~\ref{tab:mc_unobserved_variables} shows that the estimator recovers the target elasticities accurately in this benchmark. Because the missing variable is constructed from the same relationship that generated the data, the two panels produce the same estimates. Appendix~\ref{app:mc_value_error} relaxes this benchmark by allowing the value measure supplied to the estimator to deviate from the zero-profit-implied object.

\begin{table}[htbp]
\centering
\caption{Monte Carlo Results with Constructed Variables}
\label{tab:mc_unobserved_variables}
\begin{tabular}{l ccc ccc}
    \toprule
     & \multicolumn{3}{c}{$\widehat{\bar{D}}_k$} 
     & \multicolumn{3}{c}{$\widehat{\bar{D}}_l$} \\
     \cmidrule(lr){2-4} \cmidrule(lr){5-7}
    DGP & Target & Mean & RMSE & Target & Mean & RMSE \\
    \midrule
    \multicolumn{7}{l}{\textit{Panel A: Unobserved capital}} \\
    CD, CRS   & 0.600 & 0.600 & 0.001 & 0.400 & 0.398 & 0.032 \\
    CD, DRS   & 0.510 & 0.510 & 0.001 & 0.340 & 0.337 & 0.032 \\
    CES, CRS  & 0.442 & 0.442 & 0.001 & 0.558 & 0.559 & 0.032 \\
    CES, DRS  & 0.181 & 0.181 & 0.000 & 0.669 & 0.671 & 0.031 \\[6pt]
    \multicolumn{7}{l}{\textit{Panel B: Unobserved housing value}} \\
    CD, CRS   & 0.600 & 0.600 & 0.001 & 0.400 & 0.398 & 0.032 \\
    CD, DRS   & 0.510 & 0.510 & 0.001 & 0.340 & 0.337 & 0.032 \\
    CES, CRS  & 0.442 & 0.442 & 0.001 & 0.558 & 0.559 & 0.032 \\
    CES, DRS  & 0.181 & 0.181 & 0.000 & 0.669 & 0.671 & 0.031 \\
    \bottomrule
\end{tabular}
\begin{minipage}{0.9\textwidth}\scriptsize
    \textit{Notes:} 500 Monte Carlo replications with one observed construction per region-period, $J=1{,}000$ regions, and $T=5$ periods. Panel~A reports results when capital input $K$ is treated as unobserved; Panel~B reports results when housing value $V$ is treated as unobserved. In both panels, the missing variable is recovered from the same zero-profit accounting relationship used in the data-generating process. The data-generating process uses the same technology and latent-state parameters as the baseline Monte Carlo simulation. ``Target'' denotes the cross-replication mean of the realized true average elasticity. ``Mean'' denotes the mean point estimate across replications. RMSE is computed from the replication-level deviation between the estimate and the realized true value, allowing for heterogeneous CES elasticities across observations.
\end{minipage}
\end{table}

\subsection{Deviations from the zero-profit-implied value object}\label{app:mc_value_error}

The preceding subsection considers a benchmark in which the missing variable is constructed from the same zero-profit accounting relationship used in the data-generating process. This subsection relaxes that benchmark by supplying the estimator with a perturbed value measure. The exercise is motivated by the empirical application, where housing value is not directly observed and the value measure used in estimation is the zero-profit-implied object, $V_{jt}^{zp}=K_{jt}+R_{jt}$. This object is a feasible revenue measure rather than an observed transaction value. Table~\ref{tab:mcs_value_error} reports how the estimated elasticities change when the estimator observes a value measure that deviates from $V_{jt}^{zp}$, holding the production technology and latent-state process fixed.

We introduce deviations by replacing the value supplied to the estimator with $V_{jt}^{obs}=V_{jt}^{zp}\exp(m_{jt})$. The benchmark sets $m_{jt}=0$. The i.i.d.\ markup specification adds an observation-level multiplicative lognormal deviation, normalized so that $\mathbb{E}[\exp(m_{jt})]=1$. The region markup specification adds a persistent region-level multiplicative deviation, also normalized to have mean one. The remaining specifications make the deviation systematic by correlating $m_{jt}$ with either land prices or the capital-land ratio. These exercises are not tests of the zero-profit condition. They quantify the sensitivity of the estimator to deviations between the constructed value measure and the revenue object required by the model.

Three patterns emerge. First, random observation-level and persistent region-level deviations leave the mean elasticity estimates close to the targets, although coverage for the capital elasticity can deteriorate even when the average bias is small. Second, deviations systematically related to land prices or capital intensity are more consequential. This is most visible in the CES DRS design, where the land elasticity is biased downward under both systematic deviations. Third, the pattern is consistent with the role of the value measure in the two-step estimator. The perturbed value enters both the capital share in Step~1 and the adjusted value in Step~2. Systematic deviations can therefore affect the recovered capital component and can also be loaded onto the land-specific integration component $C(l)$, whereas random deviations mainly add noise and are less likely to shift average elasticity estimates.

\begin{table}[htbp]
\centering
\caption{Monte Carlo Results with Perturbed Value Measures}
\label{tab:mcs_value_error}
\small
\setlength{\tabcolsep}{3pt}
\begin{tabular}{l rrrr rrrr}
\toprule
 & \multicolumn{4}{c}{$\widehat{\bar{D}}_k$} 
 & \multicolumn{4}{c}{$\widehat{\bar{D}}_l$} \\
\cmidrule(lr){2-5} \cmidrule(lr){6-9}
Value deviation 
& Target & Mean & Bias & 95\% Cov. 
& Target & Mean & Bias & 95\% Cov. \\
\midrule
\multicolumn{9}{l}{\textit{Panel A: Cobb--Douglas, CRS}} \\
None             & 0.600 & 0.600 & 0.000 & 0.95 & 0.400 & 0.401 & 0.001 & 0.95 \\
i.i.d.\ markup   & 0.600 & 0.603 & 0.003 & 0.33 & 0.400 & 0.398 & $-$0.002 & 0.95 \\
Region markup    & 0.600 & 0.603 & 0.003 & 0.70 & 0.400 & 0.397 & $-$0.003 & 0.95 \\
Land-price corr. & 0.600 & 0.603 & 0.003 & 0.11 & 0.400 & 0.398 & $-$0.002 & 0.95 \\
K/L corr.        & 0.600 & 0.603 & 0.003 & 0.06 & 0.400 & 0.398 & $-$0.002 & 0.94 \\[5pt]

\multicolumn{9}{l}{\textit{Panel B: CES, CRS}} \\
None             & 0.442 & 0.442 & 0.000 & 1.00 & 0.558 & 0.559 & 0.001 & 0.95 \\
i.i.d.\ markup   & 0.442 & 0.444 & 0.002 & 1.00 & 0.558 & 0.557 & $-$0.001 & 0.94 \\
Region markup    & 0.442 & 0.444 & 0.002 & 0.98 & 0.558 & 0.557 & $-$0.001 & 0.96 \\
Land-price corr. & 0.442 & 0.435 & $-$0.007 & 0.00 & 0.558 & 0.566 & 0.008 & 0.95 \\
K/L corr.        & 0.442 & 0.435 & $-$0.007 & 0.00 & 0.558 & 0.567 & 0.009 & 0.94 \\[5pt]

\multicolumn{9}{l}{\textit{Panel C: CES, DRS}} \\
None             & 0.181 & 0.181 & 0.000 & 1.00 & 0.669 & 0.670 & 0.001 & 0.95 \\
i.i.d.\ markup   & 0.181 & 0.182 & 0.001 & 1.00 & 0.669 & 0.670 & 0.001 & 0.94 \\
Region markup    & 0.181 & 0.182 & 0.001 & 0.97 & 0.669 & 0.669 & 0.000 & 0.95 \\
Land-price corr. & 0.181 & 0.179 & $-$0.002 & 0.01 & 0.669 & 0.608 & $-$0.061 & 0.74 \\
K/L corr.        & 0.181 & 0.178 & $-$0.003 & 0.00 & 0.669 & 0.602 & $-$0.067 & 0.58 \\
\bottomrule
\end{tabular}
\begin{minipage}{0.95\textwidth}\scriptsize
\textit{Notes:} 500 Monte Carlo replications. The DGP is reported in the panel headers. The estimator observes the perturbed value measure $V_{jt}^{obs}=V_{jt}^{zp}\exp(m_{jt})$, while the underlying production data-generating process is unchanged. Each replication has one observed construction per region-period, $J=1{,}000$ regions, and $T=5$ periods. ``i.i.d.\ markup'' adds an observation-level multiplicative lognormal deviation normalized so that $\mathbb{E}[\exp(m_{jt})]=1$. ``Region markup'' adds a persistent region-level multiplicative lognormal deviation, also normalized to have mean one. ``Land-price corr.'' and ``K/L corr.'' introduce deviations correlated with log land prices and the log capital-land ratio, respectively. Target is the cross-replication mean of the realized true average elasticity. Mean is the average point estimate across replications. Bias is Mean minus Target. 95\% Cov.\ is the fraction of replications in which the target elasticity lies within $\widehat{\theta}_r \pm 1.96\,SD(\widehat{\theta})$, using the simulation standard deviation within each row.
\end{minipage}
\end{table}

\subsection{Quasi-panel and pseudo-panel constructions}\label{app:mc_quasi_pseudo}

Repeated cross-sectional construction data rarely provide a true panel because the same parcel is seldom rebuilt in consecutive periods. We therefore compare two ways of implementing the lagged-state moments. A \textit{quasi-panel} uses lagged input proxies at the observed-construction level. In the empirical application, these proxies are constructed by matching each current observation to nearby previous-period observations using building coordinates. In this Monte Carlo exercise, the quasi-panel is an idealized benchmark in which the relevant lagged observation is cleanly linked across periods. A \textit{pseudo-panel}, by contrast, aggregates observations to location-time cells and uses lagged cell averages in the Step~2 moment conditions. Because the pseudo-panel averages over within-cell variation, it can be less informative than the quasi-panel for recovering the land-specific component. This subsection examines this trade-off, with $N$ denoting the number of observed constructions per location-time cell.

Table~\ref{tab:lag_comparison} reports the comparison for $N \in \{2,3,5,10\}$. Both specifications are approximately unbiased across the Cobb--Douglas and CES designs. The main difference is sampling variability. Relative to the idealized quasi-panel benchmark, the pseudo-panel has substantially larger standard deviations, and the gap widens as $N$ increases. The ratio $\mathrm{SD}_{\text{pseudo}}/\mathrm{SD}_{\text{quasi}}$ rises from about $1.6$ at $N=2$ to roughly $6$ at $N=10$ in the Cobb--Douglas design. This pattern indicates that averaging over broader cells dilutes the lagged-input variation used in the Step~2 moments. The capital elasticity $\widehat{\bar{D}}_k$ is nearly identical across the two constructions and is therefore omitted from the table.

\begin{table}[htbp]
\centering
\caption{Quasi-panel vs.~pseudo-panel (Land Elasticity $\widehat{\bar{D}}_l$)}
\label{tab:lag_comparison}
\small
\begin{tabular}{ccccc}
\toprule
 & \multicolumn{2}{c}{Quasi-panel} & \multicolumn{2}{c}{Pseudo-panel} \\
\cmidrule(lr){2-3} \cmidrule(lr){4-5}
$N$ & Bias & SD & Bias & SD \\
\midrule
\multicolumn{5}{l}{\textit{Panel A: Cobb--Douglas, CRS (True $\bar{D}_l = 0.400$)}} \\
2  & $-$0.0013 & 0.0223 & $-$0.0006 & 0.0365 \\
3  &    0.0034 & 0.0222 &    0.0085 & 0.0437 \\
5  &    0.0001 & 0.0148 &    0.0078 & 0.0509 \\
10 &    0.0017 & 0.0107 &    0.0091 & 0.0672 \\[5pt]
\multicolumn{5}{l}{\textit{Panel B: CES ($\sigma=2$), CRS (True $\bar{D}_l \approx 0.558$)}} \\
2  & $-$0.0075 & 0.0224 & $-$0.0093 & 0.0393 \\
3  &    0.0000 & 0.0126 &    0.0014 & 0.0323 \\
5  &    0.0012 & 0.0128 &    0.0083 & 0.0427 \\
10 & $-$0.0008 & 0.0110 & $-$0.0135 & 0.0562 \\
\bottomrule
\end{tabular}
\begin{minipage}{0.9\textwidth}\scriptsize
    \textit{Notes:} 50 Monte Carlo replications. $J=1{,}000$ regions and $T=5$ periods. $N$ denotes the number of observed constructions per region-period. The quasi-panel uses observed-construction-level lagged inputs in the Step~2 moment conditions; the pseudo-panel uses region-year lagged averages of the same variables. The quasi-panel is an idealized benchmark rather than a simulation of geographic nearest-neighbor mismatch. The capital elasticity $\widehat{\bar{D}}_k$ is omitted because it is virtually identical across specifications.
\end{minipage}
\end{table}

\subsection{Spatial correlation in the latent state}\label{app:mc_spatial}
Table~\ref{tab:mcs_sar} reports Monte Carlo results when the latent state is spatially correlated across regions. The initial latent state is drawn from the spatial autoregressive process $(I-\rho W)^{-1}\varepsilon$, with $\rho=0.5$ and $K=6$ nearest-neighbor weights. Subsequent periods follow an AR(1) process with $\delta_1=0.9$, with the innovation variance matched to the baseline design.

The proposed estimator continues to recover the target elasticities accurately across all four designs. The mean estimates of $\widehat{\bar{D}}_k$ coincide with the targets to three decimal places, the RMSE remains small for both capital and land elasticities, and the Monte Carlo coverage rates are close to the nominal 95\% rate. These results indicate that the baseline Monte Carlo conclusions are not driven by an independent cross-sectional latent-state design. They should be interpreted as finite-sample evidence under this spatially correlated DGP, not as a general proof that the maintained assumptions hold under all forms of spatial dependence.

\begin{table}[htbp]
\centering
\caption{Monte Carlo Results with Spatially Correlated Latent States}
\label{tab:mcs_sar}
\small
\begin{tabular}{lrrrrrrrr}
\toprule
 & \multicolumn{4}{c}{$\widehat{\bar{D}}_k$}
 & \multicolumn{4}{c}{$\widehat{\bar{D}}_l$} \\
\cmidrule(lr){2-5} \cmidrule(lr){6-9}
 & Target & Mean & RMSE & 95\% Cov & Target & Mean & RMSE & 95\% Cov \\
\midrule
\multicolumn{9}{l}{\textit{Panel A: Cobb--Douglas}} \\
CRS & 0.600 & 0.600 & 0.001 & 0.95 & 0.400 & 0.402 & 0.021 & 0.95 \\
DRS & 0.510 & 0.510 & 0.001 & 0.96 & 0.340 & 0.341 & 0.020 & 0.95 \\[5pt]
\multicolumn{9}{l}{\textit{Panel B: CES}} \\
CRS & 0.448 & 0.448 & 0.001 & 1.00 & 0.552 & 0.553 & 0.018 & 0.95 \\
DRS & 0.183 & 0.183 & 0.000 & 1.00 & 0.667 & 0.667 & 0.019 & 0.95 \\
\bottomrule
\end{tabular}
\begin{minipage}{0.9\textwidth}\scriptsize
    \textit{Notes:} 500 Monte Carlo replications. The DGP is reported in the panel headers. The initial latent state is drawn from the spatial autoregressive process $(I-\rho W)^{-1}\varepsilon$, with $\rho=0.5$ and $K=6$ nearest-neighbor weights. Subsequent periods follow an AR(1) process with $\delta_1=0.9$, with the innovation variance matched to the baseline design. Each replication has one observed construction per region-period, $J=1{,}000$ regions, and $T=5$ periods. The quasi-panel specification is used for Step~2 lags. Under CES, output elasticities vary across observations, so the reported targets are cross-replication means of the realized true average elasticities. Mean denotes the average point estimate across replications. RMSE is computed from the replication-level deviation between the estimate and the realized true value. 95\% Cov is the Monte Carlo coverage rate, computed as the fraction of replications in which the target elasticity lies within $\hat{\theta}_r \pm 1.96 \times SD(\hat{\theta})$. CRS = constant returns to scale; DRS = decreasing returns to scale with $\nu=0.85$.
\end{minipage}
\end{table}

\subsection{Departures from the scalar-state and Markov restrictions}
\label{app:mc_assumption_violations}

The baseline estimator maintains a scalar revenue-relevant state and a first-order Markov law of motion. Table~\ref{tab:mcs_assumption_violations} reports sensitivity exercises that perturb these restrictions, using the CES CRS design as the common benchmark. These exercises are not tests of the maintained assumptions. They show how the estimator behaves in calibrated environments where the scalar-state or Markov restrictions are only approximate.

The baseline row leaves the benchmark DGP unchanged. The omitted-persistent-state design multiplies housing value by an additional persistent region-time revenue component that is not included in the maintained scalar-state specification. The selection design changes the observed sample by retaining observations with probabilities increasing in capital intensity and land price. It changes the target elasticity because the retained observations have a different distribution of input choices. It also reduces the effective Step~2 sample size: the Markov moments require both a current observation and a matched previous-period observation, so observation-level selection can break quasi-panel lag links. In the simulations, the average number of valid Step~2 observations falls from about $4{,}000$ in the baseline design to about $1{,}107$. The second-order-state design targets the Markov restriction. It adds a revenue component that depends on two lags, so the lagged scalar state is no longer sufficient for forecasting the current revenue state.

The estimator is stable in the omitted-persistent-state and second-order-state designs in this calibration. The capital-elasticity estimates remain close to the targets, while the land-elasticity RMSE increases moderately. The selection design is more demanding because it changes both the effective sample and the target elasticity. Even in that design, the capital elasticity remains close to the target, while the land elasticity becomes less precise. These results suggest that, in these simulations, departures from the maintained state structure mainly affect the Step~2 land-elasticity recovery rather than the Step~1 share-equation recovery of the capital elasticity.

\begin{table}[htbp]
\centering
\caption{Monte Carlo Robustness to Assumption Violations}
\label{tab:mcs_assumption_violations}
\small
\setlength{\tabcolsep}{3pt}
\begin{tabular}{lrrrcrrrc}
\toprule
 & \multicolumn{4}{c}{$\widehat{\bar{D}}_k$} & \multicolumn{4}{c}{$\widehat{\bar{D}}_l$} \\
\cmidrule(lr){2-5} \cmidrule(lr){6-9}
Design & Target\ & Mean & RMSE & 95\% Cov. & Target\ & Mean & RMSE & 95\% Cov. \\
\midrule
Baseline & 0.442 & 0.442 & 0.001 & 1.00 & 0.558 & 0.554 & 0.031 & 0.95 \\
Omitted persistent state & 0.442 & 0.442 & 0.001 & 1.00 & 0.558 & 0.554 & 0.031 & 0.95 \\
Selection on intensity & 0.488 & 0.484 & 0.004 & 0.99 & 0.512 & 0.515 & 0.066 & 0.95 \\
Second-order state & 0.442 & 0.442 & 0.002 & 1.00 & 0.558 & 0.556 & 0.040 & 0.95 \\
\bottomrule
\end{tabular}
\begin{minipage}{0.92\textwidth}\scriptsize
\textit{Notes:} 300 Monte Carlo replications. The DGP is CES with constant returns to scale, and $N=1$, $J=1{,}000$, and $T=5$. Under CES the output elasticities vary across observations, so the reported ``Target'' values are cross-simulation means. Est.\ denotes the mean point estimate across replications. RMSE is computed as the simulation-by-simulation deviation between the estimate and the realized target value. 95\% Cov.\ is the fraction of replications in which the target elasticity lies within $\hat{\theta}_r \pm 1.96\,SD(\hat{\theta})$, using the simulation standard deviation within each row.
\end{minipage}
\end{table}

\subsection{Constant physical productivity with output-price variation}
\label{app:mc_constant_productivity_price}

Table~\ref{tab:mcs_constant_productivity_price_cdg_gnr} reports an auxiliary Monte Carlo exercise in which physical productivity is fixed at a common constant for all observations, while output-price variation follows the baseline DGP. This exercise separates physical productivity from the local price component of the revenue state. It asks whether the CDG benchmark and the proposed estimator recover the production elasticities when capital choices respond to output-price variation rather than to productivity variation.

The results show that both estimators perform well in the Cobb--Douglas designs. In the CES designs, the proposed estimator recovers the target capital elasticity, while the CDG benchmark shows a small upward deviation in the CRS case. The exercise therefore indicates that the transmission problem can arise from revenue-relevant local variation even when physical productivity itself is held fixed.

\begin{table}[htbp]
\centering
\caption{Monte Carlo Results with Constant Productivity}
\label{tab:mcs_constant_productivity_price_cdg_gnr}
\small
\begin{tabular}{lccccc}
\toprule
 & Elasticity & Target & Mean & Bias & SD \\
\midrule
\multicolumn{6}{l}{\textit{Panel A: Cobb--Douglas, CRS}} \\
CDG      & $\widehat{\bar{D}}_k$ & 0.600 & 0.598 & $-$0.002 & 0.001 \\
Proposed & $\widehat{\bar{D}}_k$ & 0.600 & 0.600 & 0.000 & 0.001 \\
 & $\widehat{\bar{D}}_l$ & 0.400 & 0.401 & 0.001 & 0.015 \\[6pt]
\multicolumn{6}{l}{\textit{Panel B: Cobb--Douglas, DRS}} \\
CDG      & $\widehat{\bar{D}}_k$ & 0.510 & 0.508 & $-$0.002 & 0.001 \\
Proposed & $\widehat{\bar{D}}_k$ & 0.510 & 0.510 & 0.000 & 0.000 \\
 & $\widehat{\bar{D}}_l$ & 0.340 & 0.341 & 0.001 & 0.015 \\[6pt]
\multicolumn{6}{l}{\textit{Panel C: CES, CRS}} \\
CDG      & $\widehat{\bar{D}}_k$ & 0.435 & 0.440 & 0.005 & 0.001 \\
Proposed & $\widehat{\bar{D}}_k$ & 0.435 & 0.435 & 0.000 & 0.001 \\
 & $\widehat{\bar{D}}_l$ & 0.565 & 0.566 & 0.001 & 0.016 \\[6pt]
\multicolumn{6}{l}{\textit{Panel D: CES, DRS}} \\
CDG      & $\widehat{\bar{D}}_k$ & 0.179 & 0.179 & 0.000 & 0.000 \\
Proposed & $\widehat{\bar{D}}_k$ & 0.179 & 0.179 & 0.000 & 0.000 \\
 & $\widehat{\bar{D}}_l$ & 0.671 & 0.672 & 0.001 & 0.017 \\
\bottomrule
\end{tabular}
\begin{minipage}{0.95\textwidth}\scriptsize
\textit{Notes:} Results are based on 500 Monte Carlo replications. Physical productivity is fixed at a common constant, and output-price variation follows the baseline DGP. CD DGP: $\beta_k=0.6$, $\beta_l=0.4$. CES DGP: $\beta_k=0.4$, $\beta_l=0.6$, $\sigma=2$. DRS: $\nu=0.85$. Each replication has $N=3$ observed constructions per region-period, $J=1{,}000$ regions, and $T=5$ periods. ``Proposed'' denotes the GNR-based estimator. CDG denotes the integration-based estimator of \citet{Combes2021-pu}. The proposed estimator uses construction-level lagged links in Step~2.
\end{minipage}
\end{table}

\subsection{Simulation results for Epple, Gordon, and Sieg (2010)}
\label{app:mc_egs}

Table~\ref{tab:egs_variants} reports simulation results for several simple EGS-style specifications. For each specification, the table reports the benchmark value obtained when $\sigma_\varepsilon=0$, the Monte Carlo mean when $\sigma_\varepsilon=0.1$, and the difference between the two. The exercise is intended to show how these regression-based implementations behave under ex post output shocks and under alternative production technologies.

Under Cobb--Douglas, the specifications remain close to their benchmark values, although the log-linear slope and the implied capital share show larger positive deviations than the level and quadratic specifications. This pattern is consistent with the sensitivity of log transformations to ex post output shocks.

Under CES, the benchmark values should be interpreted as pseudo-true projection coefficients because the exact CES price--value relationship is nonlinear. The log-linear, linear, and quadratic specifications therefore do not directly recover structural CES cost-share parameters. Table~\ref{tab:egs_variants} is consistent with this interpretation: when $\sigma_\varepsilon=0.1$, the quadratic specification remains close to its CES benchmark, while the log-linear and linear specifications exhibit larger changes.

These results should be interpreted as diagnostics for simple EGS-style implementations rather than as a general critique of the EGS identification argument. The table shows that, even under CRS, functional-form approximation and ex post output shocks can affect the recovered coefficients when the underlying technology is not Cobb--Douglas.

\begin{table}[htbp]
\centering
\caption{Monte Carlo Diagnostics for Epple, Gordon, and Sieg (2010) Specifications}
\label{tab:egs_variants}
\scriptsize
\begin{tabular}{lcccccc}
\toprule
 & \multicolumn{3}{c}{CD, CRS ($\beta_k=0.6$)} 
 & \multicolumn{3}{c}{CES, CRS ($\sigma=2$, $\beta_k=0.4$)} \\
\cmidrule(lr){2-4}\cmidrule(lr){5-7}
Specification & Benchmark & Mean & Dev. & Benchmark & Mean & Dev. \\
\midrule
Log-linear slope $\widehat{b}$ $\;(\to 1/\sigma)$ 
& 1.000 & 1.040 & $+0.040$ & 0.676 & 0.697 & $+0.021$ \\
Log-linear $\widehat{\alpha}_K=1-\exp(\widehat{a})$ $\;(\to \alpha_K)$ 
& 0.600 & 0.633 & $+0.033$ & 0.198 & 0.224 & $+0.026$ \\
Linear slope $\widehat{b}$, through origin $\;(\to \alpha_L)$ 
& 0.400 & 0.409 & $+0.009$ & 0.339 & 0.352 & $+0.013$ \\
Quadratic $\widehat{b}_1$, through origin 
& 0.400 & 0.392 & $-0.008$ & 0.467 & 0.468 & $+0.001$ \\
Quadratic $\widehat{b}_2$, through origin 
& 0.000 & 0.001 & $+0.001$ & $-$0.005 & $-$0.005 & $+0.000$ \\
\bottomrule
\end{tabular}
\begin{minipage}{0.9\textwidth}\scriptsize
\textit{Notes:} 100 Monte Carlo replications. The Benchmark column reports the Monte Carlo mean when $\sigma_\varepsilon=0$. The Mean column reports the Monte Carlo mean when $\sigma_\varepsilon=0.1$. Each replication has one observed construction per region-period, $J=1{,}000$ regions, and $T=5$ periods. Each row reports a different EGS-style regression parameter. The log-linear slope $\widehat{b}$ maps to $1/\sigma$, the log-linear intercept gives $\widehat{\alpha}_K=1-\exp(\widehat{a})$, the linear-in-levels slope estimates the land cost share $\alpha_L$, and the quadratic specification reports the two coefficients of $P^L=\widehat{b}_1 V^L+\widehat{b}_2(V^L)^2$. Under CD and CRS, $P^L=\beta_l V^L$ is exact, so the benchmark values coincide with the structural values. Under CES, the implicit relation $(\beta_k/\beta_l)^\sigma(P^L)^\sigma+P^L=V^L$ has no closed-form log-linear, linear, or quadratic representation; the reported benchmark values are pseudo-true projection targets at $\sigma_\varepsilon=0$. Dev.\ $=$ Mean $-$ Benchmark.
\end{minipage}
\end{table}

\setcounter{table}{0}
\renewcommand{\thetable}{D\arabic{table}}
\setcounter{figure}{0}
\renewcommand{\thefigure}{D\arabic{figure}}
\setcounter{equation}{0}
\renewcommand{\theequation}{D\arabic{equation}}
\section{Additional results of empirical analysis}\label{app:empirical_robustness}

\subsection{Matching and sample selection}
\label{app:data_matching}

This subsection documents the matching procedure used to construct the empirical estimation sample. Table~\ref{tab:matching_summary} reports the number of observations retained at each stage. Table~\ref{tab:match_quality} summarizes the quality of the retained matches along the continuous dimensions used in the matching algorithm. Table~\ref{tab:matching_selection} compares the final matched sample with the unmatched and full-source samples. The matched sample is not fully representative of either source population; the empirical estimates should therefore be interpreted as applying to the matched new-construction sample in Tokyo's 23 special wards, rather than to all housing starts or all parcels in Tokyo.

\begin{table}[htbp]
\centering
\caption{Matching Summary}
\label{tab:matching_summary}
\begin{tabular}{lc}
\toprule
 & Survey on Building Construction Started \\
\midrule
Total observations & 62,082 \\
Eligible observations for matching & 62,082 \\
Unique matches & 27,902 \\
Match rate (\%) & 44.94 \\
Final estimation sample after trimming & 23,144 \\
Final usable share (\%) & 37.28 \\
\bottomrule
\end{tabular}
\begin{minipage}{0.9\textwidth}\scriptsize
    \textit{Notes:} This table reports matching counts from the perspective of the Survey on Building Construction Started. All 62,082 observations in Tokyo's 23 special wards over 2012--2016 are eligible for matching. The matching procedure yields 27,902 unique matches. After trimming, the final estimation sample contains 23,144 observations. The final usable share is computed relative to the total number of observations in the Survey on Building Construction Started.
\end{minipage}
\end{table}

\begin{table}[htbp]
\centering
\caption{Match Quality}
\label{tab:match_quality}
\begin{tabular}{lcccc}
\toprule
 & Mean & Median & P90 & P95 \\
\midrule
Floor area absolute difference (m$^2$) & 12.2 & 1.7 & 23.9 & 60.9 \\
Floor area percentage difference (\%) & 2.27 & 0.82 & 6.55 & 10.59 \\
Lot size absolute difference (m$^2$) & 9.0 & 0.0 & 13.1 & 72.6 \\
Lot size percentage difference (\%) & 2.91 & 0.00 & 6.03 & 28.90 \\
Composite score & 5.18 & 1.02 & 15.72 & 31.17 \\
\bottomrule
\end{tabular}
\begin{minipage}{0.9\textwidth}\scriptsize
    \textit{Notes:} The table reports matching quality for the final matched estimation sample ($N=23{,}144$). By construction, ward code, number of stories, housing type, and structure type are matched exactly. The continuous variables are matched using floor area and lot size. The composite score is the nearest-neighbor matching criterion based on the continuous matching dimensions, with lower values indicating closer matches. The median percentage difference is 0.82\% for floor area and 0.00\% for lot size.
\end{minipage}
\end{table}

\begin{table}[htbp]
\centering
\caption{Sample Selection into the Matched Sample}
\label{tab:matching_selection}
\begin{tabular}{lcc}
\toprule
 & Matched sample & Comparison sample \\
\midrule
\multicolumn{3}{l}{\textit{Panel A. Matched vs. unmatched parcels in Current Land Use}} \\
Apartment share (\%) & 56.0 & 13.4 \\
Non-wood share (\%) & 90.4 & 64.5 \\
Mean lot size (m$^2$) & 265 & 512 \\
\addlinespace
\multicolumn{3}{l}{\textit{Panel B. Matched sample vs. full Building Construction Started sample}} \\
Apartment share (\%) & 56.0 & 27.7 \\
Non-wood share (\%) & 90.4 & 33.6 \\
\bottomrule
\end{tabular}
\begin{minipage}{0.9\textwidth}\scriptsize
    \textit{Notes:} The matched sample refers to the final matched estimation sample. Panel A compares matched observations with unmatched parcels in the Survey on Current Land Use. Panel B compares the matched sample with the full set of observations in the Survey on Building Construction Started. The matched sample over-represents apartment and non-wood structures relative to both comparison groups.
\end{minipage}
\end{table}

\subsection{Estimated series coefficients and transition diagnostics}

Table~\ref{tab:gnr_params} reports the finite-dimensional coefficients used in the proposed estimator for the matched estimation sample. These coefficients are reported for transparency and reproducibility; the main empirical objects are the implied elasticities reported in Table~\ref{tab:estimates}. The Step~1 coefficients show that the fitted capital elasticity varies with the input mix. Around the centered sample mean, the positive coefficient on $k$ and the negative coefficient on $l$ imply a higher fitted capital elasticity for more capital-intensive observations. The quadratic terms indicate curvature in the fitted elasticity function. In Step~2, the coefficients of $C(l)$ determine the land-specific integration component used to recover the land elasticity. The intercept $\alpha_0$ is a level normalization and should not be interpreted as a separate structural parameter.

Table~\ref{tab:gnr_transition_params} reports a post-estimation diagnostic for the recovered revenue state. The auxiliary transition regression is not a production-function coefficient table. It summarizes the persistence and nonlinearity of the recovered state under the normalization used in Step~2.

\begin{table}[htbp]
\centering
\caption{Step 1 and Step 2 Coefficients of the Proposed Method}
\label{tab:gnr_params}
\small
\begin{tabular}{lcccc}
\toprule
 & \multicolumn{2}{c}{Full Matched Sample} & \multicolumn{2}{c}{Single-Family} \\
\cmidrule(lr){2-3} \cmidrule(lr){4-5}
 & Est. & SE & Est. & SE \\
\hline
\multicolumn{5}{l}{\textit{Step 1 Capital Elasticity Function $D_k(k,l)$}} \\[3pt]
$g_0$ (constant) & 0.493 & 0.013 & 0.420 & 0.014 \\
$g_k$ ($k$) & 0.216 & 0.007 & 0.209 & 0.011 \\
$g_l$ ($l$) & $-$0.226 & 0.006 & $-$0.216 & 0.005 \\
$g_{kk}$ ($k^2$) & 0.024 & 0.002 & 0.024 & 0.004 \\
$g_{kl}$ ($k \cdot l$) & $-$0.049 & 0.005 & $-$0.061 & 0.004 \\
$g_{ll}$ ($l^2$) & 0.023 & 0.003 & 0.026 & 0.003 \\
\\
\multicolumn{5}{l}{\textit{Step 2 GMM parameters}} \\[3pt]
$\alpha_0$ & 18.581 & 0.053 & 18.367 & 0.056 \\
$\alpha_l$ & 0.490 & 0.011 & 0.566 & 0.013 \\
$\alpha_{ll}$ & 0.118 & 0.005 & 0.111 & 0.006 \\
\bottomrule
\end{tabular}
\begin{minipage}{0.9\textwidth}\scriptsize
    \textit{Notes:} Step~1 estimates the share equation $\ln s_{jt}=\ln D_k(k_{jt},l_{jt})-\epsilon_{jt}$ by nonlinear least squares, where $s_{jt}=K_{jt}/V_{jt}$ and $D_k(k,l)=g_0+g_k k+g_l l+g_{kk}k^2+g_{kl}kl+g_{ll}l^2$. Step~2 estimates $C(l)=\alpha_0+\alpha_l l+\alpha_{ll}l^2$ by GMM. The implied land elasticity is $D_l(k,l)=g_l k+\frac{g_{kl}}{2}k^2+2g_{ll}lk+\alpha_l+2\alpha_{ll}l$. The variables $k$ and $l$ denote demeaned log capital and demeaned log land. The intercept $\alpha_0$ is included for numerical normalization and is not separately identified from the mean of the recovered revenue state. SEs are ward-cluster bootstrap SEs with $B=300$.
\end{minipage}
\end{table}

\begin{table}[htbp]
\centering
\caption{Post-Estimation Check for the Markov Process}
\label{tab:gnr_transition_params}
\small
\begin{tabular}{lcccc}
\toprule
 & \multicolumn{2}{c}{Full Matched Sample} & \multicolumn{2}{c}{Single-Family} \\
\cmidrule(lr){2-3} \cmidrule(lr){4-5}
 & Est. & SE & Est. & SE \\
\midrule
$\delta_0$ & $-$0.0048 & 0.0014 & $-$0.0055 & 0.0028 \\
$\delta_1$ & 0.8786 & 0.0065 & 0.8678 & 0.0108 \\
$\delta_2$ & 0.0320 & 0.0180 & 0.0785 & 0.0271 \\
$\delta_3$ & $-$0.1204 & 0.0132 & $-$0.1532 & 0.0192 \\
\bottomrule
\end{tabular}
\begin{minipage}{0.9\textwidth}\scriptsize
    \textit{Notes:} This table is a post-estimation diagnostic for the Markov restriction, not a production-function coefficient table. The auxiliary regression is $\hat{\omega}_{jt} = \delta_0 + \delta_1\hat{\omega}_{m(j,t)} +\delta_2\hat{\omega}_{m(j,t)}^2 + \delta_3\hat{\omega}_{m(j,t)}^3 +u_{jt}$, where $\hat{\omega}_{m(j,t)}$ is the matched previous-period state proxy. Auxiliary SEs are conditional OLS standard errors for the transition regression and are not the bootstrap SEs used for the elasticity estimates.
\end{minipage}
\end{table}

\subsection{Inference robustness}
\label{app:inference_robustness}

Table~\ref{tab:inference_robustness} compares bootstrap standard errors for the proposed-method elasticity estimates under alternative resampling schemes. The point estimates are unchanged across rows; only the bootstrap resampling scheme changes. The i.i.d.\ bootstrap treats observations as independent, the census-cell bootstrap resamples at the code3 census-cell level, and the ward bootstrap resamples at the ward level.

Ward clustering yields the largest standard errors in both samples. In the full matched sample, the ward-cluster standard error for $\widehat{\bar{D}}_k$ is about an order of magnitude larger than the i.i.d.\ bootstrap standard error, and the ward-cluster standard error for $\widehat{\bar{D}}_l$ is roughly three times as large. We therefore use ward-cluster bootstrap standard errors in the main empirical tables as a conservative baseline.

\begin{table}[htbp]
\centering
\caption{Inference Robustness for the Proposed-Method Elasticities}
\label{tab:inference_robustness}
\small
\begin{tabular}{lcccccc}
\toprule
\multirow{2}{*}{Bootstrap scheme}
& \multicolumn{3}{c}{Matched sample}
& \multicolumn{3}{c}{Single-family sample} \\
\cmidrule(lr){2-4} \cmidrule(lr){5-7}
& $\widehat{\bar{D}}_k$ & $\widehat{\bar{D}}_l$ & RTS
& $\widehat{\bar{D}}_k$ & $\widehat{\bar{D}}_l$ & RTS \\
\midrule
i.i.d.
& 0.510 & 0.486 & 0.996
& 0.440 & 0.563 & 1.004 \\
& $(0.001)$ & $(0.003)$ & $(0.003)$
& $(0.002)$ & $(0.006)$ & $(0.006)$ \\[2pt]

Census-cell cluster
& 0.510 & 0.486 & 0.996
& 0.440 & 0.563 & 1.004 \\
& $(0.003)$ & $(0.005)$ & $(0.003)$
& $(0.004)$ & $(0.007)$ & $(0.005)$ \\[2pt]

Ward cluster
& 0.510 & 0.486 & 0.996
& 0.440 & 0.563 & 1.004 \\
& $(0.012)$ & $(0.012)$ & $(0.004)$
& $(0.012)$ & $(0.012)$ & $(0.006)$ \\
\bottomrule
\end{tabular}
\begin{minipage}{0.92\textwidth}\scriptsize
    \textit{Notes:} The table reports proposed-method elasticity estimates with bootstrap standard errors in parentheses. All entries use 300 bootstrap replications. Ward clustering resamples Tokyo's 23 special wards. RTS denotes returns to scale and is computed as $\widehat{\bar{D}}_k+\widehat{\bar{D}}_l$ within each bootstrap replication.
\end{minipage}
\end{table}

\subsection{Common-support comparison with the CDG benchmark}
\label{app:cdg_support}

Table~\ref{tab:estimates_cdg_support} repeats the main elasticity comparison on the central input support used for the CDG smoothing grid. This check asks whether the comparison between the proposed estimator and the CDG benchmark is driven by differences in the input support over which the estimates are evaluated. The sample keeps observations with land and capital between their 10th and 90th percentiles. The estimates remain close to the baseline results. In both the full matched sample and the single-family subsample, the proposed estimator yields larger capital elasticities than the CDG benchmark, and the implied returns to scale remain close to one.

\begin{table}[htbp]
\centering
\caption{Production Function Estimates on the CDG Support}
\label{tab:estimates_cdg_support}
\small
\begin{tabular}{lcccccc}
\toprule
 & \multicolumn{3}{c}{Full Sample} & \multicolumn{3}{c}{Single-Family} \\
 \cmidrule(lr){2-4} \cmidrule(lr){5-7}
 & Estimate & SE & 95\% CI & Estimate & SE & 95\% CI \\
\hline
\multicolumn{7}{l}{\textit{Panel A: CDG (Combes, Duranton, Gobillon 2021)}} \\[3pt]
$\widehat{\bar{D}}_k$ & 0.452 & 0.014 & $[0.425,\ 0.480]$ & 0.337 & 0.013 & $[0.312,\ 0.362]$ \\
\\
\multicolumn{7}{l}{\textit{Panel B: Proposed method on CDG support}} \\[3pt]
$\widehat{\bar{D}}_k$ & 0.520 & 0.014 & $[0.493,\ 0.548]$ & 0.456 & 0.014 & $[0.429,\ 0.483]$ \\
$\widehat{\bar{D}}_l$ & 0.478 & 0.013 & $[0.452,\ 0.504]$ & 0.539 & 0.014 & $[0.512,\ 0.566]$ \\
$\widehat{\bar{D}}_k + \widehat{\bar{D}}_l$ & 0.998 & 0.005 & $[0.988,\ 1.009]$ & 0.995 & 0.007 & $[0.981,\ 1.008]$ \\
\bottomrule
\end{tabular}
\begin{minipage}{0.9\textwidth}\scriptsize
    \textit{Notes:} The CDG-support sample keeps observations with land and capital between their 10th and 90th percentiles, matching the support used to evaluate the CDG smoothing grid. Full matched sample: $N=15{,}300$; single-family subsample: $N=6{,}618$. SEs are ward-cluster bootstrap SEs with $B=300$. 95\% CI = estimate $\pm$ 1.96 $\times$ SE. All estimates use year-demeaned variables following \citet{Combes2021-pu}. Panel~B uses one-nearest-neighbor quasi-panel lags constructed from building centroid coordinates. The returns-to-scale SE is computed as the bootstrap standard deviation of $\widehat{\bar{D}}_k+\widehat{\bar{D}}_l$.
\end{minipage}
\end{table}

\subsection{Elasticity heterogeneity}
\label{app:elasticity_heterogeneity}

\begin{figure}[htbp]
\centering
\includegraphics[width=0.78\textwidth]{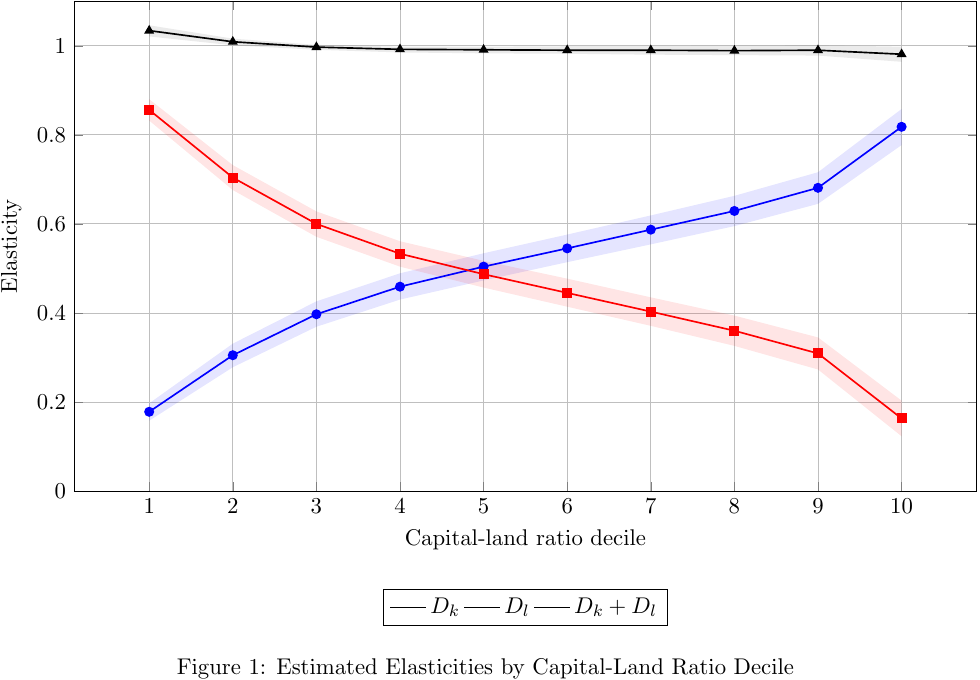}
\caption{Elasticity Heterogeneity by Capital-Land Ratio Decile}
\label{fig:elasticity_by_kl_decile_appendix}
\begin{minipage}{0.86\textwidth}\scriptsize
\textit{Notes:} The figure reports average observation-level elasticities by deciles of the capital-land ratio. Shaded bands report 95\% confidence intervals based on ward-cluster bootstrap standard errors with 300 replications. The deciles are computed in the matched estimation sample.
\end{minipage}
\end{figure}

The average elasticities in Table~\ref{tab:estimates} summarize heterogeneous observation-level elasticities. This subsection reports how the recovered elasticities vary across observable groups. Table~\ref{tab:elasticity_heterogeneity} reports group-specific averages by housing type and land price quartile. Figure~\ref{fig:elasticity_by_kl_decile_appendix} reports the corresponding pattern by capital-land ratio decile, which is the most direct way to visualize how the proposed estimator allows elasticities to vary with the input mix. These exercises are descriptive and should not be interpreted as separate identification exercises.

\begin{table}[htbp]
\centering
\caption{Elasticity Heterogeneity by Housing Type and Land Price}
\label{tab:elasticity_heterogeneity}
\small
\begin{tabular}{llrrrr}
\toprule
Margin & Group & $N$ & $\widehat{\bar{D}}_k$ & $\widehat{\bar{D}}_l$ & RTS \\
\midrule
\multicolumn{6}{l}{\textit{Panel A. Housing type}} \\
Housing type & Non-single-family & 12{,}954 & 0.563 & 0.427 & 0.990 \\
& & & $(0.015)$ & $(0.015)$ & $(0.004)$ \\
Housing type & Single-family & 10{,}190 & 0.443 & 0.561 & 1.004 \\
& & & $(0.010)$ & $(0.010)$ & $(0.003)$ \\
\addlinespace
\multicolumn{6}{l}{\textit{Panel B. Land price quartile}} \\
Land price & Q1 & 5{,}792 & 0.473 & 0.526 & 0.999 \\
& & & $(0.014)$ & $(0.014)$ & $(0.003)$ \\
Land price & Q2 & 5{,}787 & 0.502 & 0.495 & 0.996 \\
& & & $(0.015)$ & $(0.015)$ & $(0.003)$ \\
Land price & Q3 & 5{,}786 & 0.520 & 0.476 & 0.995 \\
& & & $(0.018)$ & $(0.018)$ & $(0.004)$ \\
Land price & Q4 & 5{,}779 & 0.547 & 0.447 & 0.994 \\
& & & $(0.020)$ & $(0.019)$ & $(0.004)$ \\
\bottomrule
\end{tabular}
\begin{minipage}{0.92\textwidth}\scriptsize
\textit{Notes:} The table reports group-specific averages of observation-level elasticities from the proposed estimator. Ward-cluster bootstrap standard errors based on 300 replications are in parentheses. Land price groups are sample quartiles. RTS denotes returns to scale and is computed as $\widehat{\bar{D}}_k+\widehat{\bar{D}}_l$ within each group.
\end{minipage}
\end{table}

\subsection{Value construction sensitivity}
\label{app:value_sensitivity}

Table~\ref{tab:value_sensitivity} evaluates the sensitivity of the estimates to the maintained zero-profit-implied value object, $V^{zp}=K+R$. Because housing value is not directly observed in the empirical application, the table asks how the recovered elasticities change when the value measure supplied to the estimator is perturbed. The proportional rows scale the entire constructed value measure, while the remaining rows change the relative weight placed on land cost or construction cost. These exercises are not tests of the zero-profit condition. They show how the estimated elasticities move under alternative constructed value measures.

The results show that the capital-land split is sensitive to the value construction. Uniformly scaling up the value measure lowers the recovered capital elasticity, while increasing the weight on land cost shifts the estimates toward a lower capital elasticity and a higher land elasticity. The returns-to-scale estimates move less than the individual input elasticities, but they also decline when the constructed value is scaled up substantially. Thus, the returns-to-scale estimates are less sensitive than the allocation between capital and land elasticities across these perturbations, but they remain conditional on the maintained value construction.

\begin{table}[htbp]
\centering
\caption{Sensitivity to the Constructed Value Measure}
\label{tab:value_sensitivity}
\small
\begin{tabular}{lcccccc}
\toprule
\multirow{2}{*}{Value construction}
& \multicolumn{3}{c}{Matched sample}
& \multicolumn{3}{c}{Single-family sample} \\
\cmidrule(lr){2-4} \cmidrule(lr){5-7}
& $\widehat{\bar{D}}_k$ & $\widehat{\bar{D}}_l$ & RTS
& $\widehat{\bar{D}}_k$ & $\widehat{\bar{D}}_l$ & RTS \\
\midrule
Baseline: $K+R$ & 0.510 & 0.486 & 0.996 & 0.440 & 0.563 & 1.004 \\
& $(0.012)$ & $(0.012)$ & $(0.003)$ & $(0.013)$ & $(0.013)$ & $(0.005)$ \\
$1.05(K+R)$ & 0.486 & 0.500 & 0.986 & 0.419 & 0.569 & 0.988 \\
& $(0.012)$ & $(0.011)$ & $(0.004)$ & $(0.011)$ & $(0.012)$ & $(0.005)$ \\
$1.10(K+R)$ & 0.464 & 0.514 & 0.978 & 0.400 & 0.574 & 0.974 \\
& $(0.010)$ & $(0.010)$ & $(0.004)$ & $(0.012)$ & $(0.013)$ & $(0.006)$ \\
$1.20(K+R)$ & 0.425 & 0.537 & 0.962 & 0.367 & 0.583 & 0.950 \\
& $(0.010)$ & $(0.010)$ & $(0.005)$ & $(0.011)$ & $(0.013)$ & $(0.006)$ \\
$K+0.8R$ & 0.558 & 0.438 & 0.995 & 0.487 & 0.515 & 1.002 \\
& $(0.012)$ & $(0.011)$ & $(0.004)$ & $(0.012)$ & $(0.012)$ & $(0.005)$ \\
$K+1.2R$ & 0.471 & 0.526 & 0.997 & 0.402 & 0.603 & 1.005 \\
& $(0.012)$ & $(0.012)$ & $(0.004)$ & $(0.012)$ & $(0.012)$ & $(0.006)$ \\
$0.9K+R$ & 0.542 & 0.477 & 1.019 & 0.465 & 0.575 & 1.039 \\
& $(0.013)$ & $(0.012)$ & $(0.004)$ & $(0.013)$ & $(0.012)$ & $(0.005)$ \\
$1.1K+R$ & 0.482 & 0.494 & 0.976 & 0.418 & 0.554 & 0.972 \\
& $(0.010)$ & $(0.010)$ & $(0.004)$ & $(0.011)$ & $(0.012)$ & $(0.006)$ \\
\bottomrule
\end{tabular}
\begin{minipage}{0.92\textwidth}\scriptsize
\textit{Notes:} The baseline value construction is $V^{zp}=K+R$, where $K$ is construction cost and $R$ is approximated land acquisition cost. Ward-cluster bootstrap standard errors based on 300 replications are in parentheses. The proportional rows scale the entire constructed value measure. The remaining rows perturb either the land-cost component or the construction-cost component. The table reports sample-average elasticities from the proposed estimator under each value construction. RTS denotes returns to scale and is computed as $\widehat{\bar{D}}_k+\widehat{\bar{D}}_l$.
\end{minipage}
\end{table}

\subsection{Lag construction and placebo checks}
\label{app:lag_checks}

The results show that the baseline estimates are stable across feasible local lag constructions. In Table~\ref{tab:pseudo_panel_lag_checks}, Panel~A shows the expected pattern: as the administrative cells become coarser, the land-elasticity estimates become less precise and move farther from the baseline. Small census-cell and sub-ward averages remain close to the nearest-neighbor baseline, whereas ward-level averages generate larger deviations and much larger standard errors. This pattern is consistent with the idea that broad spatial aggregation dilutes the local lag information used in Step~2.

Table~\ref{tab:quasi_panel_lag_checks} shows that one-to-one geographic matches are stable when the previous-period match is restricted to nearby observations. The threshold rows remain close to the baseline, even when the match is required to be within 50 or 100 meters. The placebo rows provide a sharper diagnostic. Random previous-year and wrong-period lag assignments produce much larger standard errors and unstable land-elasticity estimates, showing that the Step~2 estimates are not invariant to arbitrary lag assignments. The capital elasticity is unchanged across these checks because it is identified in Step~1 and does not use the Markov lag construction.

\begin{table}[htbp]
\centering
\caption{Pseudo-Panel and Averaged-Lag Robustness}
\label{tab:pseudo_panel_lag_checks}
\small
\begin{tabular}{lrrrrrrr}
\toprule
\multirow{2}{*}{Lag construction}
& \multirow{2}{*}{Avg. dist.}
& \multicolumn{3}{c}{Matched sample}
& \multicolumn{3}{c}{Single-family sample} \\
\cmidrule(lr){3-5} \cmidrule(lr){6-8}
& & $\widehat{\bar{D}}_k$ & $\widehat{\bar{D}}_l$ & RTS
& $\widehat{\bar{D}}_k$ & $\widehat{\bar{D}}_l$ & RTS \\
\midrule
Baseline: nearest neighbor & 149.9 & 0.510 & 0.486 & 0.996 & 0.440 & 0.563 & 1.004 \\
& & $(0.012)$ & $(0.012)$ & $(0.003)$ & $(0.013)$ & $(0.013)$ & $(0.005)$ \\
\addlinespace
\multicolumn{8}{l}{\textit{Panel A. Administrative-cell averages}} \\
Small census cells & -- & 0.510 & 0.479 & 0.989 & 0.440 & 0.559 & 0.999 \\
& & $(0.013)$ & $(0.010)$ & $(0.005)$ & $(0.012)$ & $(0.011)$ & $(0.007)$ \\
Sub-ward census cells & -- & 0.510 & 0.494 & 1.004 & 0.440 & 0.555 & 0.995 \\
& & $(0.012)$ & $(0.012)$ & $(0.004)$ & $(0.013)$ & $(0.014)$ & $(0.008)$ \\
Ward cells & -- & 0.510 & 0.533 & 1.044 & 0.440 & 0.622 & 1.063 \\
& & $(0.012)$ & $(0.074)$ & $(0.070)$ & $(0.013)$ & $(0.071)$ & $(0.070)$ \\
\addlinespace
\multicolumn{8}{l}{\textit{Panel B. Geographic averaged lags}} \\
KNN average, $K=3$ & 223.7 & 0.510 & 0.492 & 1.002 & 0.440 & 0.570 & 1.011 \\
& & $(0.012)$ & $(0.014)$ & $(0.005)$ & $(0.012)$ & $(0.017)$ & $(0.009)$ \\
KNN average, $K=5$ & 280.4 & 0.510 & 0.494 & 1.004 & 0.440 & 0.569 & 1.010 \\
& & $(0.012)$ & $(0.015)$ & $(0.006)$ & $(0.012)$ & $(0.018)$ & $(0.009)$ \\
Radius average within 100m & 63.6 & 0.510 & 0.488 & 0.998 & 0.440 & 0.565 & 1.006 \\
& & $(0.012)$ & $(0.012)$ & $(0.005)$ & $(0.012)$ & $(0.015)$ & $(0.009)$ \\
Radius average within 300m & 193.4 & 0.510 & 0.484 & 0.994 & 0.440 & 0.568 & 1.008 \\
& & $(0.011)$ & $(0.010)$ & $(0.004)$ & $(0.012)$ & $(0.012)$ & $(0.007)$ \\
\bottomrule
\end{tabular}
\begin{minipage}{0.94\textwidth}\scriptsize
    \textit{Notes:} The baseline row reproduces the proposed-method estimates in Table~\ref{tab:estimates}. Ward-cluster bootstrap standard errors based on 300 replications are in parentheses. Administrative-cell averages aggregate previous-year observations to location-year cells and use lagged cell averages in Step~2. Small census cells are the finest observed census cells in the estimation data. Sub-ward census cells aggregate those cells to a coarser within-ward geography. Ward cells aggregate to Tokyo's 23 special wards. Geographic averaged lags use building coordinates but average over multiple previous-year observations near each current observation. KNN average selects the $K$ nearest previous-year observations and averages their lagged variables. Radius average uses all previous-year observations within the stated radius. Average distance is reported in meters when applicable. The capital elasticity is identified in Step~1 and therefore does not vary with the Step~2 lag construction. RTS denotes returns to scale and is computed as $\widehat{\bar{D}}_k+\widehat{\bar{D}}_l$.
\end{minipage}
\end{table}

\begin{table}[htbp]
\centering
\caption{One-to-One Geographic Quasi-Panel and Placebo Lag Checks}
\label{tab:quasi_panel_lag_checks}
\small
\begin{tabular}{lrrrrrrr}
\toprule
\multirow{2}{*}{Lag construction}
& \multirow{2}{*}{Avg. dist.}
& \multicolumn{3}{c}{Matched sample}
& \multicolumn{3}{c}{Single-family sample} \\
\cmidrule(lr){3-5} \cmidrule(lr){6-8}
& & $\widehat{\bar{D}}_k$ & $\widehat{\bar{D}}_l$ & RTS
& $\widehat{\bar{D}}_k$ & $\widehat{\bar{D}}_l$ & RTS \\
\midrule
Baseline: nearest neighbor & 149.9 & 0.510 & 0.486 & 0.996 & 0.440 & 0.563 & 1.004 \\
& & $(0.012)$ & $(0.012)$ & $(0.003)$ & $(0.013)$ & $(0.013)$ & $(0.005)$ \\
\addlinespace
\multicolumn{8}{l}{\textit{Panel A. One-to-one geographic quasi-panel matches}} \\
Nearest within 50m & 31.3 & 0.510 & 0.496 & 1.006 & 0.440 & 0.574 & 1.014 \\
& & $(0.013)$ & $(0.014)$ & $(0.006)$ & $(0.012)$ & $(0.016)$ & $(0.010)$ \\
Nearest within 100m & 59.2 & 0.510 & 0.488 & 0.998 & 0.440 & 0.563 & 1.003 \\
& & $(0.011)$ & $(0.011)$ & $(0.005)$ & $(0.013)$ & $(0.015)$ & $(0.009)$ \\
Nearest within 200m & 102.2 & 0.510 & 0.487 & 0.997 & 0.440 & 0.567 & 1.007 \\
& & $(0.012)$ & $(0.011)$ & $(0.004)$ & $(0.013)$ & $(0.013)$ & $(0.008)$ \\
Nearest within 300m & 126.6 & 0.510 & 0.486 & 0.996 & 0.440 & 0.564 & 1.004 \\
& & $(0.012)$ & $(0.011)$ & $(0.004)$ & $(0.012)$ & $(0.012)$ & $(0.006)$ \\
\addlinespace
\multicolumn{8}{l}{\textit{Panel B. Placebo lag assignments}} \\
Random previous-year match & -- & 0.510 & 0.567 & 1.078 & 0.440 & 0.886 & 1.326 \\
& & $(0.012)$ & $(0.302)$ & $(0.302)$ & $(0.012)$ & $(0.297)$ & $(0.296)$ \\
Random next-year match & -- & 0.510 & 0.235 & 0.745 & 0.440 & 0.954 & 1.394 \\
& & $(0.012)$ & $(0.283)$ & $(0.299)$ & $(0.012)$ & $(0.299)$ & $(0.314)$ \\
\bottomrule
\end{tabular}
\begin{minipage}{0.94\textwidth}\scriptsize
    \textit{Notes:} The baseline row reproduces the proposed-method estimates in Table~\ref{tab:estimates}. Ward-cluster bootstrap standard errors based on 300 replications are in parentheses. One-to-one geographic quasi-panel rows match each current observation to one previous-year observation using building centroid coordinates. The threshold rows retain only observations with a previous-year nearest neighbor inside the stated distance. The placebo rows deliberately assign nonlocal or wrong-period lags and therefore break the maintained local lag structure. Average distance is reported in meters. The capital elasticity is identified in Step~1 and therefore does not vary with the Step~2 lag construction. RTS denotes returns to scale and is computed as $\widehat{\bar{D}}_k+\widehat{\bar{D}}_l$.
\end{minipage}
\end{table}

\subsection{EGS results}
\label{app:egs_empirical}

Table~\ref{tab:egs_empirical} reports EGS-style estimates in the matched estimation sample under three functional forms. The purpose of this exercise is diagnostic. It summarizes the empirical relationship between land prices and housing value per unit of land, but the structural interpretation of the coefficients requires the EGS maintained assumptions, including constant returns to scale and the relevant duality mapping.

The estimates show a positive relationship between land price and value per unit of land across specifications. In the log-linear specification, the estimated slope is $\widehat{b}=0.5369$ with a ward-cluster bootstrap standard error of $0.0705$. In the linear specification, the slope is $\widehat{b}=0.4224$ with a standard error of $0.0257$. In the quadratic specification, the first-order term is $\widehat{b}_1=0.3511$, while the second-order term is small in magnitude, $\widehat{b}_2=4.49\times 10^{-8}$.

The structural mapping from these reduced-form coefficients is sensitive to the functional form. For example, under the log-linear Cobb--Douglas interpretation, the slope implies $\widehat{\sigma}=1/\widehat{b}=1.8627$, but the intercept would imply a capital share outside the economically meaningful range. We therefore interpret the EGS estimates as reduced-form diagnostics rather than as structural estimates of the housing production function in the Tokyo data.

\begin{table}[htbp]
\centering
\caption{EGS-Style Estimates}
\label{tab:egs_empirical}
\small
\begin{tabular}{llcc}
\toprule
Specification & Parameter & Estimate & SE \\
\midrule
\multicolumn{4}{l}{\textit{Log-linear}} \\
& $\widehat{a}$ & 5.3632 & 0.9265 \\
& $\widehat{b}$ & 0.5369 & 0.0705 \\
\addlinespace
\multicolumn{4}{l}{\textit{Linear}} \\
& $\widehat{b}$ & 0.4224 & 0.0257 \\
\addlinespace
\multicolumn{4}{l}{\textit{Quadratic}} \\
& $\widehat{b}_1$ & 0.3511 & 0.0274 \\
& $\widehat{b}_2$ & $4.49\times 10^{-8}$ & $1.84\times 10^{-8}$ \\
\bottomrule
\end{tabular}
\begin{minipage}{0.9\textwidth}\scriptsize
    \textit{Notes:} The effective sample size is $N=23{,}144$. Standard errors are ward-cluster bootstrap standard errors based on 300 bootstrap replications. The log-linear specification regresses log land price per unit of land on log housing value per unit of land. The linear and quadratic specifications use the corresponding level relationship between land price per unit of land and housing value per unit of land. The estimates are reported as EGS-style reduced-form diagnostics; their structural interpretation requires the maintained EGS assumptions.
\end{minipage}
\end{table}

\clearpage
\begin{spacing}{1.0}
\ifx\undefined\bysame
\newcommand{\bysame}{\leavevmode\hbox to\leftmargin{\hrulefill\,\,}}
\fi

\end{spacing}

\end{document}